\documentclass[11pt]{iopart}

\expandafter\let\csname equation*\endcsname\relax
\expandafter\let\csname endequation*\endcsname\relax 

\usepackage[english]{babel}
\usepackage{amsmath}
\usepackage{amssymb}
\usepackage{graphicx}
\usepackage{color}
\usepackage[utf8]{inputenc}
\usepackage{amsthm}
\usepackage{placeins}
\usepackage{verbatim}
\usepackage{booktabs}
\usepackage{color}
\usepackage{bbold}
\usepackage{comment}
\usepackage{url}
\usepackage[all]{xy}
\usepackage[width=17.5cm, vscale=0.8]{geometry}
\newcommand{\bm}{{\bf m}}
\newcommand{\be}{\begin{equation}}
\newcommand{\ee}{\end{equation}}
\newcommand{\bea}{\begin{eqnarray}}
\newcommand{\eea}{\end{eqnarray}}

\newcommand{\order}{{\mathcal{O}}}
\newcommand{\bra}{{\langle}}
\newcommand{\ket}{{\rangle}}
\newcommand{\rd}{{\rm d}}
\newcommand{\bx}{{\bf x}}

\usepackage{epsfig}
\newcommand{\bxi}{{\mbox{\boldmath $\xi$}}}
\newcommand{\bsigma}{{\mbox{\boldmath $\sigma$}}}

\begin{document}
\title{Associative Networks with Diluted Patterns:
Dynamical Analysis at Low and Medium Load
}

\author{S Bartolucci$^{1}$, A Annibale $^{1,2}$}

\address{$^1$ Department of Mathematics, King's College London, The Strand,
London WC2R 2LS, UK}
\address{$^2$ Institute for Mathematical and Molecular Biomedicine, King's College London, Hodgkin Building, London SE1 1UL, UK}
\date{\today}

\begin{abstract}
In this work we solve the dynamics of pattern diluted associative networks,
evolving via sequential Glauber update. We derive dynamical equations for the 
order parameters, that quantify the simultaneous pattern recall of the system, 
and analyse the nature and stability of the stationary solutions by means of 
linear stability analysis as well as Monte Carlo simulations. 
We investigate the parallel retrieval capabilities of the 
system in different regions of the phase space, in particular in the low and 
medium storage regimes and for finite and extreme pattern dilution.
Results show that in the absence of patterns cross-talk, all patterns 
are recalled symmetrically for any temperature below 
criticality, while in the presence of pattern cross-talk, symmetric 
retrieval becomes unstable as temperature is lowered and 
a hierarchical retrieval takes over. The shape of the hierarchical retrieval occurring at zero temperature is provided. The parallel retrieval capabilities of the network are seen to degrade gracefully in the regime of strong interference, but they are not destroyed. 
\end{abstract}

\section{Introduction}
Associative networks with diluted patterns have been recently introduced \cite{hierar} as a generalisation of the standard Hopfield model
\cite{hopfield,amit,hopfieldhigh}.
Interestingly, the introduction of dilution in the stored patterns of information, drastically changes the collective behaviour of the system,
switching its operational behaviour from sequential to 
parallel processing. In fact, pattern diluted associative networks are able to orchestrate a parallel retrieval of multiple patterns of information 
at the same time. 
Despite their wide range of possible applications, from artificial intelligence \cite {sollichcoolen} to theoretical immunology  \cite{PRE},
these types of memories have been studied analytically only in statics so far, where several dilution and storage regimes were investigated \cite{medium, saturation,Sollich}. 
Dynamical studies are 
limited to numerical simulations and are restricted to the low storage regime, 
where the number of stored patterns is finite \cite{stabilitya}.
 While lacking so far, the dynamical approach is generally expected to provide
 richer information than static 
analysis, which can only describe the behaviour of the system at the steady 
state, and it does not require that the system is in equilibrium.
In this work we carry out a dynamical analysis of associative memories with diluted patterns, evolving via Glauber sequential update. 
We solve the dynamics in different storage regimes, via 
non-equilibrium statistical mechanical techniques, first developed in 
\cite{coolruij,macro} for standard Hopfield networks and spin 
glasses. We derive equations for the time evolution of 
a set of macroscopic order parameters, 
quantifying the simultaneous patterns retrieval, from the dynamical laws of the microscopic neuronal system.  
We analyse the structure of the stationary solutions of the dynamics in different regimes of dilution 
(from finite to extreme), in the low and in the medium storage regimes, and we inspect
their stability, in different regions of the phase diagram,
through linear stability analysis as well as Monte Carlo simulations.
Our analysis clarifies several results from previous 
equilibrium analysis, establishing 
the limit of their validity in different regions of the tunable parameters space
(noise, dilution and storage load). Moreover, it provides new information in those regions of the 
phase space where previous results from equilibrium analysis 
correspond to unstable steady states of the dynamical equations.
As a result, this work draws a complete picture of the rich dynamical 
behaviour of pattern diluted associative networks away from saturation. 

The paper is organised as follows. In Sec.  \ref{sec:model}  we define the model and its microscopic dynamics. 
In Sec.  \ref{sec:dynamicaleq} we derive dynamical equations 
for a suitable set of macroscopic parameters, which quantify the (parallel) retrieval of the system, and we lay the basis for the analysis of the nature and 
stability of stationary solutions in different regimes of dilution and storage load. Sec.  \ref{sec:gamma0} is devoted to low storage and finite dilution,
while Sec.  \ref{sec:medium} deals with the analysis at medium load 
and extreme dilution. Finally, in Sec.  \ref{sec:conclusion} we summarise and discuss our results.

\section{The model}\label{sec:model}

We consider a system of $N$ Ising spins $\sigma_i=\pm 1$, for $i=1,\dots,N$, interacting through pairwise couplings
\begin{eqnarray}
J_{ij}= \sum_{\mu=1}^P \xi_i^{\mu}\xi_{j}^{\mu},
\label{eq:coupling}
\end{eqnarray}
where $\bxi^{\mu}=(\xi_1^\mu, \dots, \xi_N^\mu)$, with $\mu=1,...,P$, represent stored patterns of information. 
We assume that pattern entries are quenched random variables, identically and independently distributed with probability distribution
\begin{eqnarray}
P(\bxi)=\prod_{i\mu}\left[\frac{c}{2N^{\gamma}}\big(\delta_{\xi_i^{\mu},1} + \delta_{\xi_i^{\mu},-1}\big) +\bigg(1-\frac{c}{N^{\gamma}}\bigg)\delta_{\xi_i^{\mu},0}\right]
\label{eq:proba}
\end{eqnarray}
where $\gamma\in [0,1)$ and $c=\order{(N^0)}$. The number of stored patterns is $P=\alpha N^{\delta}$, with $\delta\in [0,1)$, 
$\alpha\in \mathbb{R}^+$. The fraction of non-zero entries in the stored patterns determines the degree of dilution of the system: for $\gamma=0$ the system 
is finitely diluted, whereas for $\gamma>0$ the system is extremely diluted.
For $\delta>0$ the number of patterns stored in the system diverges in the thermodynamic limit, however as long as 
$\delta<1$, the number of patterns grows sub-linearly in the system size, and the system is away from saturation.  
Spins obey random sequential Glauber dynamics via an effective local field
\begin{eqnarray}
h_i({\bsigma}) = \frac{1}{N^{1-\gamma}}\sum_{j\neq i}^NJ_{ij}\sigma_j 
\label{eq:field}
\end{eqnarray}
so that the 
probability $p_t(\bsigma)$
of finding the system in a state $\bsigma=(\sigma_1,\dots,\sigma_N)$ at time $t$, evolves according to the master 
equation
\begin{eqnarray}
\frac{\partial{p_t(\mbox{\boldmath$\sigma$}})}{\partial{t}}= \sum_{i=1}^N\big[p_t(F_i {\bsigma})w_i(F_i {\bsigma})-p_t(\mbox{\boldmath$\sigma$})w_i(\mbox{\boldmath$\sigma$})\big]
\label{eq:master}
\end{eqnarray}
where $F_i$ is the $i$-th spin-flip operator $F_i(\sigma_1,..,\sigma_i,...,\sigma_N)= (\sigma_1,...,-\sigma_i,...\sigma_N)$
and transition rates between $\bsigma$ and $F_i\bsigma$ have the Glauber form
\begin{eqnarray}
w_i(\bsigma)= \frac{1}{2}[1-\sigma_i\tanh{(\beta h_i(\bsigma))}]
\label{eq:trans}
\end{eqnarray}
where $\beta=1/T$ is the inverse temperature of the system, giving the rate of spontaneous spin flips. \\
For symmetric couplings, as those given in \eqref{eq:coupling}, the dynamics obeys detailed balance and the process evolves towards equilibrium, 
described by the Boltzmann distribution, with Hamiltonian:
\begin{eqnarray}
H= -\frac{1}{2N^{1-\gamma}} \sum_{i,j=1}^N\sum_{\mu=1}^P\xi_i^{\mu}\xi_j^{\mu}\sigma_i\sigma_j.
\label{eq:hamiltonian}
\end{eqnarray}
We note, however, that our approach does not require such restriction and it opens the way to a broader range of applications where symmetric interactions are not realistic, particularly welcome in theoretical immunology \cite{PRE}.
Up to prefactors, expression (\ref{eq:hamiltonian}) is identical to the Hamiltonian of the standard Hopfield network, but here patterns $\xi_i^{\mu}$ are diluted according to 
(\ref{eq:proba}). One can show that the prefactor provides the correct normalisation to the Hamiltonian \cite{medium}. A heuristic 
argument is that, since the number of non-zero entries in a given pattern is $\mathcal{O}(N^{1-\gamma})$, one has, for condensed (i.e. retrieved) patterns, 
$\sum_i \xi_i^\mu \sigma_i=\order{(N^{1-\gamma})}$ with the expected number of condensed patterns being  
$\mathcal{O}(N/N^{1-\gamma})=\order{(N^\gamma)}$, so their contribution to the Hamiltonian is $\order{(N)}$. Uncondensed patterns, for which 
$\sum_i \xi_i^\mu \sigma_i=\order{(N^{(1-\gamma)/2})}$, are present only for  
$\delta>\gamma$ and are $\mathcal{O}(N^{\delta})$, thus providing a 
contribution which is always subleading away from saturation, as expected.

\section{Macroscopic dynamics}\label{sec:dynamicaleq}
For large $N$, finding a solution for the set of $2^N$ coupled, non-linear 
differential equations \eqref{eq:master} is very hard, and a convenient 
approach is to use the microscopic stochastic laws for the neuronal dynamics 
\eqref{eq:master} to derive dynamical equations \eqref{eq:m}
for the probability distribution of a suitable set of macroscopic observables. As for the standard Hopfield model, it is convenient 
to choose these as (suitably normalised)
overlaps between the microscopic configurations of the system and the stored 
patterns $\bxi^\mu$, $\mu=1,\ldots,P$
\begin{eqnarray}
m_{\mu}(\bsigma)=\frac{1}{cN^{1-\gamma}}\sum_{j=1}^N\xi_j^{\mu}\sigma_j,
\label{eq:overlap}
\end{eqnarray}
with the prefactor chosen to make these order $\order{(1)}$ quantities.
These order parameters quantify the pattern retrieval of the system, as 
in the absence of retrieval they vanish in the thermodynamic limit, 
whereas an order $\order{(1)}$ pattern overlap $m_\mu$ indicates that the 
system has retrieved pattern $\bxi^\mu$. We note that for large $N$ and 
$\gamma<1$ the effective fields (\ref{eq:field}) depend on $\bsigma$ only through the overlaps, i.e.
$h_i(\bsigma)=c\sum_\mu \xi_i^\mu m_\mu(\bsigma)$.

The probability of finding the system, at time $t$, in a state where the macroscopic parameters (\ref{eq:overlap}) take values ${\bf m}= (m_1,...,m_P)$,
\begin{eqnarray}
P_t({\bf m})= \sum_{\bsigma} p_t(\mbox{\boldmath$\sigma$})\delta(\bm-\bm(\bsigma)),
\end{eqnarray}
evolves, due to equation (\ref{eq:master}) governing the microscopic probability $p_t(\bsigma)$, as
\begin{eqnarray}
\frac{\rmd P_t({\bf m})}{\rmd t}= \sum_{\bsigma}\sum_{i=1}^{N} \delta({\bf m}-{\bf m}(\bsigma))\big[p_t(F_i\bsigma)w_i(F_i\bsigma)-p_t(\bsigma)w_i(\bsigma)\big],
\end{eqnarray}
where relabelling $F_i\bsigma\to\bsigma$ in the first term on the RHS gives
\begin{eqnarray}
\frac{\rmd P_t({\bf m})}{\rmd t}= \sum_{\bsigma}\sum_{i=1}^{N} p_t(\bsigma)w_i(\bsigma)[\delta({\bf m}-{\bf m}(F_i\bsigma))-\delta({\bf m}-{\bf m}(\bsigma))].
\label{eq:master2}
\end{eqnarray}
By introducing the increments
\begin{eqnarray}
\Delta_{i,\mu}= m_{\mu}(F_i\bsigma)-m_{\mu}(\bsigma)= \frac{2}{cN^{1-\gamma}}\xi_i^{\mu} \sigma_i
\label{eq:increment}
\end{eqnarray}
and Taylor expanding (\ref{eq:master2}) in powers of $\Delta_{i,\mu}(\bsigma)$, we obtain the so-called Kramers-Moyal (KM) expansion.
In order for the expansion to hold, increments $\Delta_{i,\mu}$ must be small. This condition is satisfied in the thermodynamic limit $N\to\infty$ for 
any $\gamma<1$. The case $\gamma=1$, where each neuron is finitely connected to other neurons, violates the condition necessary to perform the KM expansion 
and will not be considered here.
By Taylor expanding the $\delta$-function in powers of $\Delta_{i\mu}$, we obtain 
\begin{eqnarray}
\hspace*{-2.5cm}\frac{d P_t({\bf m})}{d t}= \sum_{i=1}^N \sum_{\bsigma}w_i(\bsigma)p_t(\bsigma)
\sum_{\ell \geq1} \frac{(-1)^{\ell}}{\ell !}
\sum_{\mu_1=1}^P\dots\sum_{\mu_\ell=1}^P\frac{\partial^{\ell}}{\partial m_{\mu_1}\dots\partial m_{\mu_{\ell}}}
\bigg[\delta[{\bf m}-{\bf m}(\bsigma)][\Delta_{i\mu_i}\dots\Delta_{i\mu_{\ell}}]\bigg] 
\nonumber\\
\end{eqnarray}
Defining 
\begin{eqnarray}
F_{\mu_1\dots\mu_{\ell}}^{(\ell)}({\bf m},t)=\langle \sum_{i=1}^Nw_j(\bsigma)\Delta_{i\mu_i}\dots\Delta_{i\mu_{\ell}}\rangle_{{\bf m} , t}
\label{eq:F}
\end{eqnarray}
with
\begin{eqnarray}
\langle \dots \rangle_{{\bf m}, t} = \frac{\sum_{\bsigma}\delta({\bf m} - {\bf m} (\bsigma))\dots 
p_t(\bsigma)}{\sum_{\bsigma}\delta({\bf m} - {\bf m} (\bsigma))p_t(\bsigma)}
\end{eqnarray}
we can write the equation for the time evolution of $P_t(\bm)$ as
\begin{eqnarray}
\frac{\partial P_t({\bf m})}{\partial t} = \sum_{\ell \geq 1} \frac{(-1)^{\ell}}{\ell !} \sum_{\mu_1=1}^P\dots\sum_{\mu_{\ell} =1}^P\frac{\partial^\ell}{\partial m_{\mu_1}\dots\partial m_{\mu_{\ell}}}[P_t({\bf m})F_{\mu_1\dots\mu_{\ell}}^{(\ell)}({\bf m},t)]
\label{eq:KM}
\end{eqnarray}
Keeping only terms up to the first order in the $\Delta_{i\mu}$'s 
leads to the Liouville equation, corresponding to a deterministic evolution of the order parameters $\bm$. Second order terms lead to the Fokker-Planck equation, which includes diffusive processes.
Hence, a sufficient condition for the observables $\bm(\bsigma)$ to evolve deterministically, in the limit $N\to\infty$, is
\begin{eqnarray}
\lim_{N\to \infty} \sum_{\ell\geq 2}\frac{(-1)^\ell}{\ell !} \sum_{\mu_1=1}^P\dots\sum_{\mu_\ell=1}^P\sum_{i=1}^N\frac{\partial^{\ell}}{\partial m_{\mu_1}\dots\partial m_{\mu_{\ell}}}\langle |\Delta_{i\mu_i}\dots\Delta_{i\mu_{\ell}}|\rangle_{{\bf m},t}=0.
\label{eq:condition}
\end{eqnarray}
Each $\Delta_{i,\mu}$, as defined in (\ref{eq:increment}), contributes 
to the sums an order $\order{(N^{-1})}$ because patterns entries $\xi_i^\mu$ 
are non-zero with probability $cN^{-\gamma}$. Recalling
that $P=\alpha N^{\delta}$, condition (\ref{eq:condition}) reduces to
\begin{eqnarray}
\lim_{N\to \infty}N^{2\delta -1}=0
\end{eqnarray}
which is satisfied for $\delta<\frac{1}{2}$. For this range of values of the storage load parameter $\delta$, higher order terms in the KM expansion 
vanish in the thermodynamic limit and the equation for $P_t(\bm)$ reduces, 
to leading orders in $N$, to the Liouville equation 
\begin{eqnarray}
\frac{\partial P_t({\bf m})}{\partial t}= -\sum_{\mu=1}^P\frac{\partial}{\partial m_{\mu}}\left[P_t({\bf m})F_{\mu}^{(1)}({\bf m},t)\right]
\label{eq:liouville}
\end{eqnarray}
where $F_{\mu}^{(1)}$ results from insertion of (\ref{eq:trans}) and 
(\ref{eq:increment}) in (\ref{eq:F}). Setting $\hat\beta=\beta c$, we have
\begin{eqnarray}
F_{\mu}^{(1)}({\bf m},t) &=& \frac{1}{cN^{1-\gamma}}\sum_{i=1}^N\xi_i^{\mu}\tanh{(\hat{\beta} \sum_{\mu=1}^P \xi_i^\mu m_\mu)} - m_{\mu}
\nonumber\\
&=& \frac{N^{\gamma}}{c}\langle \xi^{\mu}\tanh(\hat{\beta}\sum_{\nu=1}^P\xi^{\nu}m_{\nu})\rangle_{\bxi} - m_{\mu}
\end{eqnarray}
where we denoted the average $\bra f(\bxi)\ket_{\bxi}=\sum_\bxi P(\bxi)f(\bxi)$.
We note that $F_{\mu}^{(1)}$ is $\order{(1)}$ as each $\xi^\mu$ is non-zero with probability $cN^{-\gamma}$. 
The Liouville equation corresponds to a deterministic evolution for the 
$m_\mu$'s, given by the set of ordinary, coupled, 
non-linear differential equations
\begin{eqnarray}
\frac{\rmd m_{\mu}}{\rmd t}= \frac{N^{\gamma}}{c}\langle \xi^{\mu}\tanh(\hat{\beta}\sum_{\nu=1}^P\xi^{\nu}m_{\nu})\rangle_{\bxi} - m_{\mu}.
\label{eq:m}
\end{eqnarray}
In the following sections  we will investigate the nature and stability of the 
steady state solutions of \eqref{eq:m} in different regions of the tunable parameters (noise, dilution, storage load).
\subsection{The steady state and linear stability analysis}
The steady state ($\rd \bm/dt=0$) is given by the self-consistency equation 
for $\bm$
\begin{eqnarray}
{\bf m}= \frac{N^{\gamma}}{c}\langle\bxi\tanh[\hat{\beta} ({\bf m}\cdot\bxi)]\rangle_{\bxi}.
\label{eq:vecstedy}
\end{eqnarray}
One can show that the system undergoes a phase transition at $T_c=c$, 
with the equilibrium phase at $T>T_c$ characterised by $\bm={\bf 0}$
and retrieval occurring at $T<T_c$, where $\bm\neq {\bf 0}$. 
Taking the scalar product of (\ref{eq:vecstedy}) with $\bm$ and using 
the inequality $|\tanh x|\leq |x|$, gives
\begin{eqnarray}
{\bf m}^2\leq \frac{N^{\gamma}}{c}\hat{\beta}\langle ({\bf m}\cdot\bxi)^2\rangle_{\bxi}=\hat{\beta} {\bf m}^2
\end{eqnarray}
which implies $\bm=0$ for $\hat\beta>1$. 
By linearising the system (\ref{eq:m}) about ${\bf m}=0$, 
\begin{eqnarray}
\frac{\rmd m_{\mu}}{\rmd t} = (\hat{\beta}-1)m_{\mu}
\end{eqnarray}
we find that ${\bf m}=0$ is stable for $\hat\beta<1$ and reached exponentially fast with rate $\tau=1/(1-\hat{\beta})$, 
whereas a power law decay $m_\mu\sim t^{-1/2}$ takes place for $\hat \beta \to 1$.
Hence, for $T>c$ there is no retrieval. 
On the other hand, for $T<c$ the solution $\bm=0$ becomes unstable and we expect nonzero solutions $\bm\neq 0$ to bifurcate continuously from the $\bm=0$ solution as we decrease the temperature below $T_c$.
We will inspect the stability of the non-zero steady state solutions $\bm^\star$ below $T_c$ by means of linear stability analysis, i.e. by inspecting the eigenvalues of the Jacobian of the dynamical system (\ref{eq:m}), 
\bea
A_{\mu \nu}=\frac{\partial F_{\mu}^{(1)}(\bm)}{\partial m_{\nu}}\bigg|_{\bm=\bm^\star},
\quad\quad\quad F_{\mu}^{(1)}({\bf m})=\frac{N^\gamma}{c}\langle \xi^{\mu}\tanh(\hat{\beta}
\sum_{\nu=1}^P\xi^{\nu}m_{\nu})\rangle_{\bxi} - m_{\mu}
\label{eq:Jacobian}
\eea
evaluated at the steady state $\bm^\star$. 
When the system recalls a number $n$ of patterns, we have, up to permutations of pattern indices,
$\bm^\star=(m_1,\ldots,m_n,0\ldots,0)$ with $m_\mu\neq 0~\forall~\mu\leq n$, and 
the resulting matrix
\begin{eqnarray}
A_{\mu\nu}= \frac{N^{\gamma}}{c}\hat{\beta}\langle \xi^{\mu}\xi^{\nu}(1-\tanh^2(\hat{\beta} {\bf m}\cdot \bxi)\rangle_{\bxi}-\delta_{\mu\nu}
\label{eq:matrix}
\end{eqnarray}
is a block matrix where diagonal elements are
\begin{eqnarray}
A_{\mu\mu}&=& \hat{\beta} (1-q_\mu)-1 ,\quad {\rm for~} \mu\leq n
\label{eq:Aqm}
\\
A_{\mu\mu}&=& \hat{\beta} (1-q)-1, \quad {\rm ~~for~} \mu> n
\label{eq:Aq}
\end{eqnarray}
with
\bea
q_\mu&=&\langle(\tanh^2(\hat{\beta} (m_\mu+\sum_{\nu\neq \mu}^n \xi^{\nu}m_\nu))\rangle_{\bxi}
\label{eq:qm}
\\
q&=&\langle(\tanh^2(\hat{\beta} \sum_{\nu=1}^n \xi^{\nu}m_\nu)\rangle_{\bxi},
\label{eq:q}
\end{eqnarray}
and off-diagonal elements are
\bea
A_{\mu\nu}&=& -\hat{\beta} \frac{c}{N^\gamma}Q, \quad {\rm for~}\mu,\nu\leq n
\nonumber\\
A_{\mu\nu}&=&0, \quad\quad\quad\quad {\rm ~~~otherwise}
\label{eq:Aoff}
\eea
with
\bea
Q=\langle( \xi^{\mu}\xi^{\nu}\tanh^2(\hat{\beta} {\bf m}\cdot\bxi)\rangle_{\bxi| \xi^{\mu,\nu}\neq 0},
\label{eq:off}
\eea
where the average in \eqref{eq:off} is over $\xi^{\mu,\nu}\neq 0$, so that $Q$ is an order $\order{(1)}$ quantity.

Next, we investigate the structure of the first class of solutions 
to bifurcate away from $\bm={\bf 0}$ below $T=c$. 
To this purpose, we Taylor expand (\ref{eq:vecstedy}) 
for small $|\bm|$ in powers of $\epsilon=\hat{\beta}-1$:
\begin{eqnarray}
m_{\mu}&\simeq& \frac{N^{\gamma}}{c}\bigg[\sum_{\nu=1}^P\hat{\beta}\langle \xi^{\nu}\xi^{\mu}\rangle m_{\nu}-\frac{\hat{\beta}^3}{3}\sum_{\nu,\rho,\lambda=1}^P\langle \xi^{\nu}\xi^{\mu}\xi^{\rho}\xi^{\lambda}\rangle m_{\nu}m_{\rho}m_{\lambda}
+\order{(\bm^5)}\bigg]
\nonumber\\
&=& (1+\epsilon) m_{\mu} -\frac{m_\mu}{3} (m_{\mu}^2 +\frac{3c}{N^{\gamma}}\sum_{\mu\neq\rho}^P m_{\rho}^2)+\order{(\bm^5,\epsilon\bm^3)}
\end{eqnarray}
where we have used $\hat{\beta}^3=(1+\epsilon)^3\simeq 1+3\epsilon+\ldots$ and anticipated that $m_\mu=\order{(t^{1/2})}$. This yields
\begin{eqnarray}
0= m_{\mu}\bigg[\epsilon-\frac{1}{3}\bigg(m_{\mu}^2+\frac{3c}{N^{\gamma}}({\bf m}^2 -m_{\mu}^2)\bigg)\bigg]
\end{eqnarray}
which gives either the trivial solution $m_{\mu}=0$ or 
\begin{eqnarray}
m_{\mu}^2 = \frac{3\epsilon -\frac{3c}{N^{\gamma}}{\bf m}^2}{1-\frac{3c}{N^{\gamma}}}.
\label{eq:nonzero}
\end{eqnarray}
For $\gamma>0$ and large $N$ we get
\be
m_{\mu}^2 = 3\epsilon-\frac{3c}{N^\gamma}\bm^2
\label{eq:mgnz}
\ee
whereas for $\gamma=0$ one has 
\be
m_{\mu}^2 = 3\frac{\epsilon -c \bm^2}{1-3c}
\label{eq:mgz}
\ee
Since each component $m_{\mu}$ only depends on the whole vector ${\bf m}$, it is clear that, close to criticality, we have, for each component,
$m_{\mu} \in \{-m,0,m\}$. Using the invariance of the dynamical equations 
under $m_\mu\to -m_\mu$, we can from now on focus on 
non-negative solutions, that we can write, up to permutations of 
pattern indices, as 
${\bf m} = m ( 1,\dots, 1, 0, \dots,0)$, with 
\be
{{\bf m}}^2=m^2 n,
\ee
where $n$ represents the number of non-null entries of the overlap vector. 
We will call this class of solution {\it symmetric}, as condensed patterns 
are retrieved all with the same intensity.
Different storage and dilution regimes will yield different scaling for the 
number $n$ of condensed patterns, 
as we shall detail below.
In the following sections, we will derive 
the region of stability of this class of solutions and we will discuss 
how the system breaks the symmetry at low temperature.

\subsection{Noise distribution}\label{sec:noise}
Interestingly, the nature and stability of the non-trivial solution below criticality depend on the values of the parameters $\delta, \gamma$, controlling 
the storage load and the dilution in our system, respectively.
To see this, it is convenient to work out (\ref{eq:m}), by performing the average over $\xi^\mu$,
\begin{eqnarray}
\frac{\rmd m_{\mu}}{\rmd t} = \langle \tanh(\hat{\beta}(m_{\mu} +\sum_{\nu\neq\mu}^P\xi^{\nu}m_{\nu})\rangle_{\bxi} - m_{\mu} 
\label{eq:msepar}
\end{eqnarray}
and inserting 
$1= \int_{-\infty}^{+\infty} \rmd z  \delta\big(z-\sum_{\nu\neq\mu}^P\xi^{\nu}m_{\nu}
\big)$, so we can express the dynamical equations
\be
\frac{\rmd m_{\mu}}{\rmd t}= \int_{-\infty}^{+\infty} \rmd z \tanh[\hat{\beta}(m_{\mu}+z)] P_{\mu}(z|\{m_{\nu}\}) -m_{\mu}
\label{eq:mprobz}
\ee
in terms of the noise distribution
\be
P_{\mu}(z|\{m^{\nu}\})= \langle \delta(z-\sum_{\nu\neq\mu}^P\xi^{\nu}m_{\nu})\rangle_{\bxi}.
\label{eq:Pm}
\ee
The latter can be calculated by using 
the Fourier representation of the $\delta$-function 
\begin{eqnarray}
 P_{\mu}(z |\{m_{\nu}\})=\int_{-\infty}^{+\infty} \frac{\rmd \hat z}{2\pi} \rme^{\rmi z\hat z}\langle \prod_{\nu\neq\mu}^P \rme^{-\rmi\hat{z}\xi^{\nu} m_{\nu}}\rangle_{\bxi} = \int_{-\infty}^{+\infty} \frac{\rmd \hat z}{2\pi} \rme^{\rmi z\hat z}\prod_{\nu\neq\mu}^P\langle  \rme^{-\rmi\hat{z}\xi^{\nu} m_{\nu}}\rangle_{\bxi} 
\end{eqnarray}
and performing the average over the disorder
\begin{eqnarray}
\langle \rme^{-\rmi\hat{z}\xi^{\nu} m_{\nu}}\rangle_{\xi^{\nu}} 
= 1 + \frac{c}{N^{\gamma}}(\cos(\hat{z} m_{\nu}) -1)\simeq \rme^{\frac{c}{N^{\gamma}} (\cos(\hat{z}m_{\nu})-1)}
\end{eqnarray}
where the last approximation holds for large $N$.
This leads to
\begin{eqnarray}
P_\mu(z|\{m_{\nu}\})=\int_{-\infty}^{+\infty} \frac{\rmd\hat z}{2\pi} \rme^{\rmi z\hat z} \rme^{\sum_{\nu\neq\mu}^P \frac{c}{N^{\gamma}}(\cos(\hat{z} m_{\nu})-1)}.
\label{eq:PZ}
\end{eqnarray}
Different choices of the parameters $\delta,\gamma$, 
confer different properties to the distribution (\ref{eq:PZ}), thus 
leading to different dynamical behaviour.
In particular, when $\gamma>0$, extending the sum to all $ \nu$ in (\ref{eq:PZ}) gives a negligible contribution in the thermodynamic limit, so the distribution $P_\mu(z|\{m_\nu\})$ 
converges, for large $N$, to the distribution 
\begin{eqnarray}
P(z|{\bf m})=\langle \delta(z-\sum_{\nu=1}^P m_\nu \xi^\nu) \rangle_{\bxi} 
=\int_{-\infty}^{+\infty} \frac{\rmd\hat z}{2\pi} \rme^{\rmi z\hat z+\sum_{\nu=1}^P \frac{c}{N^{\gamma}}(\cos(\hat{z} m_{\nu})-1)}
\label{eq:Pgammageq0}
\end{eqnarray}
which depends on the whole vector ${\bf m}= (m_1, \dots, m_P)$. 
This highlights the possibility of having symmetric solutions 
as fixed points of the dynamics in large regions of the phase diagram, 
meaning that all patterns are retrieved with the same intensity.
In contrast, for $\gamma=0$, (\ref{eq:PZ}) retains its dependence 
on $\mu$ in the thermodynamic limit and one expects the symmetry of patterns 
to be broken in large regions of the phase diagram, meaning that the system retrieves patterns with different 
intensities, i.e. \emph{hierarchically}.

\begin{figure}
        \centering   		
  \scalebox{.26}{\includegraphics{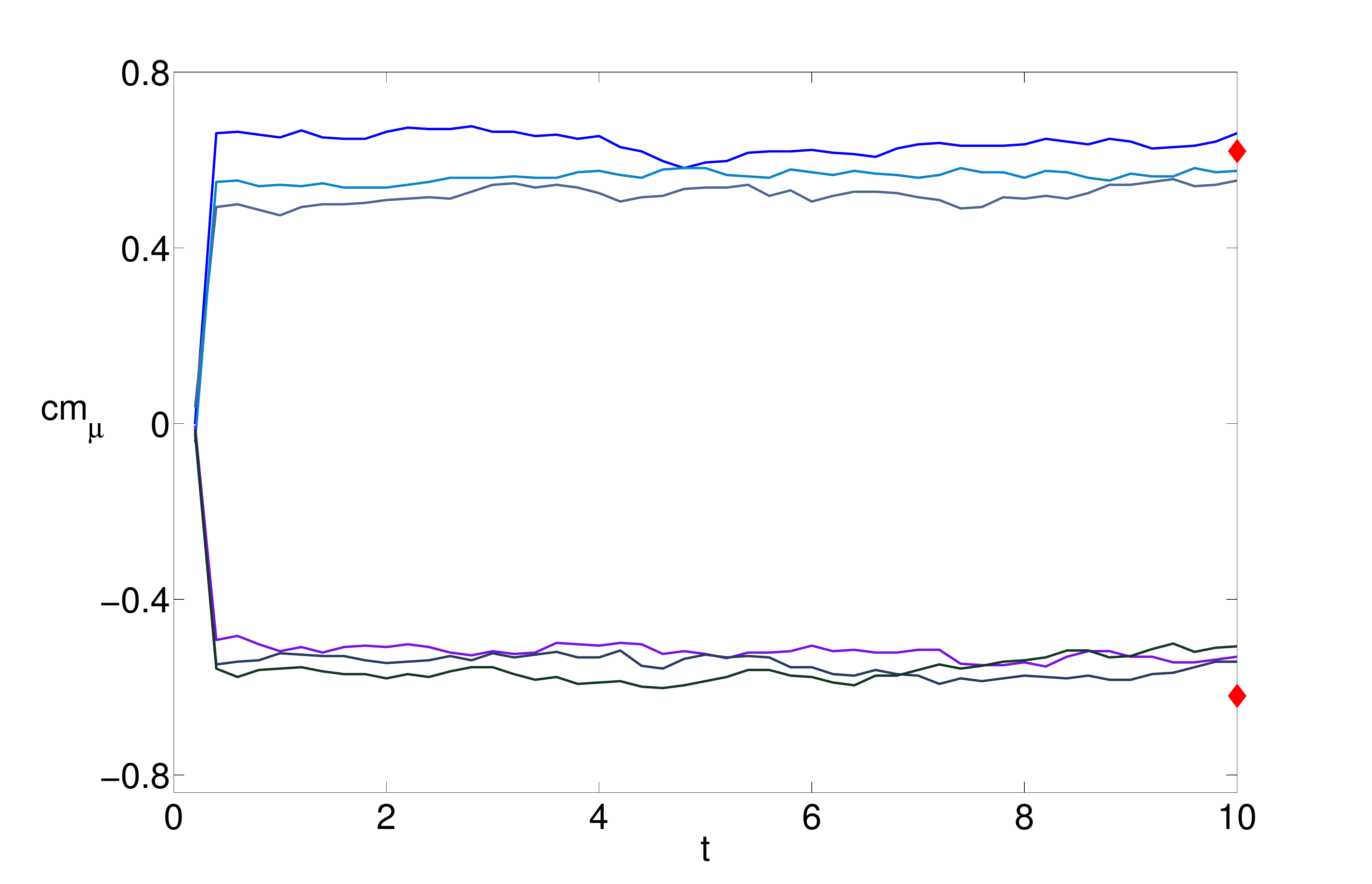}}  		  
 \caption{Results of Monte Carlo simulations with $N=10^4$ spins at $c=0.8$, $\hat{T}=T/c=0.6$, $\gamma =0.3$ and $\delta=0.25$. We plot the overlaps $m_{\mu}$ with different patterns $\bxi^{\mu}$, drawn from (\ref{eq:proba}), as a function of time $t$. Red symbols represent the magnetisations in the steady state predicted by 
(\ref{eq:decouple}). 
Deviations from the predicted values are small and compatible with finite size effects $\mathcal{O}(N^{-\gamma})$. } \label{fig:maggiore} 
        \end{figure}

In addition, we note that 
for $\gamma>\delta$, (and $\gamma>0$, since $\delta \geq 0$)
the probability distribution $P(z|{\bf m})$ reduces to a $\delta$-function in the thermodynamic limit
\begin{eqnarray}
P(z |{\bf m})=\int_{-\infty}^{+\infty} \frac{\rmd\hat z}{2\pi} \rme^{i z\hat z} \rme^{\sum_{\nu=1}^{\alpha N^{\delta}} \frac{c}{N^{\gamma}}(\cos(\hat{z} m_{\nu})-1)} \simeq \delta (z). 
\end{eqnarray} 
This leads the dynamical equations (\ref{eq:mprobz}) to uncouple and the model 
reduces to $P$ independent Hopfield models with a single stored pattern, 
or, equivalently, to a set of $P$ independent Curie-Weiss 
ferromagnets, 
\begin{eqnarray}
\frac{\rmd}{\rmd t}m_{\mu} = \tanh(\hat{\beta}m_{\mu})-m_{\mu},\quad\quad\quad \mu=1,\ldots,P
\label{eq:decouple}
\end{eqnarray}
with a rescaled critical temperature $\hat T_c=T/c$,
due to the presence of dilution in the patterns.
Each overlap approaches exponentially (one of) the non-zero (Gauge-symmetric) 
solutions of $m=\tanh(\hat\beta m)$ and 
the stable steady state of the system is given by $\bm=m(1,\ldots,1)$, 
for any temperature below criticality.
This shows that the system is able to retrieve {\it all} the stored patterns 
in parallel and that the retrieval is symmetric.
This behaviour is confirmed by Monte Carlo simulations, 
shown in figure \ref{fig:maggiore}, where the 
trajectories of the overlaps are seen to approach (up to finite size 
effects) the expected steady-states, represented by the symbols.

Conversely, for any $\gamma\leq \delta$, the dynamical equations 
(\ref{eq:mprobz}) are coupled, meaning that there is an interference between 
patterns. This holds true even when the number of patterns is finite, i.e. 
for $\delta=\gamma=0$.
In the sections below we will study the nature and stability of the 
stationary points of the dynamics for $\gamma\leq \delta$ and we 
show that despite the presence of pattern interference, the system is able to 
retrieve all stored patterns in parallel, 
although symmetric solutions will no longer be stable at all 
temperatures below criticality. 

We will analyse, separately, the dynamics of the system for $\delta=0$, i.e. the so-called {\it low storage regime} (Sec.  \ref{sec:gamma0}), where the number of stored patterns is finite, 
and for $\delta>0$, i.e. the so-called {\it medium storage regime} (Sec. \ref{sec:medium}), 
where the number of stored patterns is sub-linear in the system size.
\section{Low Storage and Finite Dilution: $\gamma=\delta=0$} \label{sec:gamma0}
We have discussed above the case $\delta<\gamma$, where overlaps of 
different patterns evolve independently and 
symmetric retrieval is stable for any temperature below criticality.  
Hence, the remainder of the paper is devoted to analyse the regime 
$\delta\geq \gamma$, where, as explained is Sec. \ref{sec:noise}, a cross-talk 
between patterns appears.
In this section we consider the case of finite 
number of patterns ($\delta=0$), and finite dilution ($\gamma=0$).
\subsection{A toy model:$P=2$}
\begin{figure}
\centering
	\includegraphics[width=0.5\textwidth]{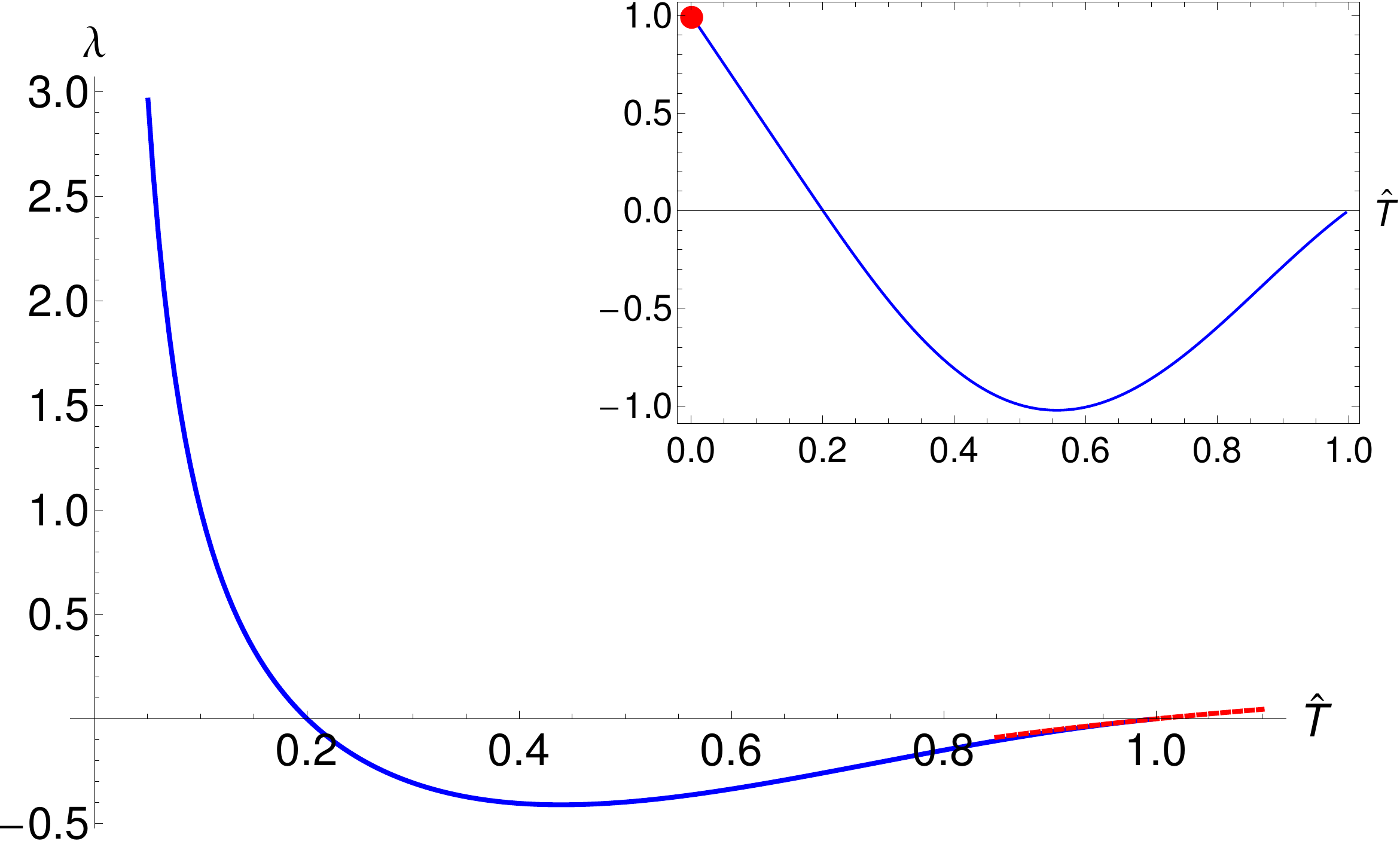}
\caption{Eigenvalue $\lambda_2$ as a function of the temperature $\hat{T}$, for a fixed $c=0.2$. The red dashed line gives the theoretical prediction near 
the critical temperature (\ref{eq:eigen2P_TC}). The figure in the inset shows agreement with (\ref{eq:eigen2P_T0}), which 
gives $\lambda_2/(\hat T c) \to 1$ as $\hat{T}\to 0$ (red marker).}
\label{fig:eigen_sec_2P}
\end{figure}

Let us first illustrate the main features of the dynamics at low storage
by considering the simple toy model with $P=2$ patterns.
For $\gamma=0$ and $P=2$ our dynamical equations (\ref{eq:m}) reduce to
\begin{eqnarray}
\hspace*{-2cm}
\frac{\rmd }{\rmd t}m_1 =f_1({\bf m})= (1-c)\tanh(\hat{\beta} m_1) +\frac{c}{2}[\tanh(\hat{\beta}(m_1+m_2)) +\tanh(\hat{\beta}(m_1-m_2)) ] -m_1
\nonumber\\
 \hspace*{-2cm}
\frac{\rmd}{ \rmd t}m_2 = f_2({\bf m})= (1-c)\tanh(\hat{\beta} m_2) +\frac{c}{2}[\tanh(\hat{\beta}(m_1+m_2)) -\tanh(\hat{\beta}(m_1-m_2) )] -m_2 
\label{eq:system2}
\end{eqnarray}
Close to criticality, we have from (\ref{eq:mgz}), $m_1=m_2=m$, with 
\begin{eqnarray}
m^2 = \frac{3\epsilon-3c\bm^2}{1-3c}
\end{eqnarray}
and $\bm^2=2m^2$, yielding
\begin{eqnarray}
m^2=\frac{3\epsilon}{1+3c}.\label{eq:symmixt}
\end{eqnarray}
We analyse the stability of the symmetric mixtures studying the eigenvalues of the Jacobian (\ref{eq:Jacobian}) evaluated at the fixed point $\bm^\star=m(1,1)$.
Solutions are linearly stable if the eigenvalues are negative.
Due to the simple form of the matrix, eigenvalues are easily found as
 \begin{eqnarray}
 \lambda_1&=&\hat{\beta}-1-(1-c)\hat{\beta}\tanh^2(\hat{\beta} m)-c\hat{\beta}\tanh^2(2\hat{\beta} m)\\
 \lambda_2&=&\hat{\beta}-1-(1-c)\hat{\beta}\tanh^2(\hat{\beta} m).
 \label{eq:eigen}
 \end{eqnarray}
We note that $\lambda_1 <\lambda_2$, hence the stability is given by the region where the largest eigenvalue $\lambda_2$ is negative.
We can work out $\lambda_2$ analytically near $T\simeq T_c$ and $T\simeq0$. 
Near the critical temperature, we expand (\ref{eq:eigen}) in powers of 
$\hat{\beta}=c\beta=1+\epsilon$ and using \eqref{eq:symmixt} we find
   \begin{eqnarray}
  \lambda_2= -\frac{2\epsilon (1-3c)}{1+3c}.
   \label{eq:eigen2P_TC}
 \end{eqnarray}
Hence, for $T\simeq T_c$ solutions are stable when $c<1/3$. 
In the opposite limit, $T\to 0$, i.e. $\beta\to \infty$, we have $\tanh^2(\beta m)\to 1$, and
 \begin{eqnarray}
 \lambda_2 \sim \hat{\beta} c-1 >0.
 \label{eq:eigen2P_T0}
 \end{eqnarray}
Hence, symmetric solutions are unstable at low temperature for any value of 
$c>\sqrt{T}$. 
The full temperature-dependence of $\lambda_2$ can be obtained numerically and is shown in figure \ref{fig:eigen_sec_2P} for
dilution $c=0.2$. The limits for temperature close to zero and to the critical point are in agreement with the theoretical predictions \eqref{eq:eigen2P_TC} and \eqref{eq:eigen2P_T0}, 
shown in the graph and in the graph inset, as the dashed line and the symbol, respectively.

\begin{figure}
\centering 
\includegraphics[width=0.4\textwidth]{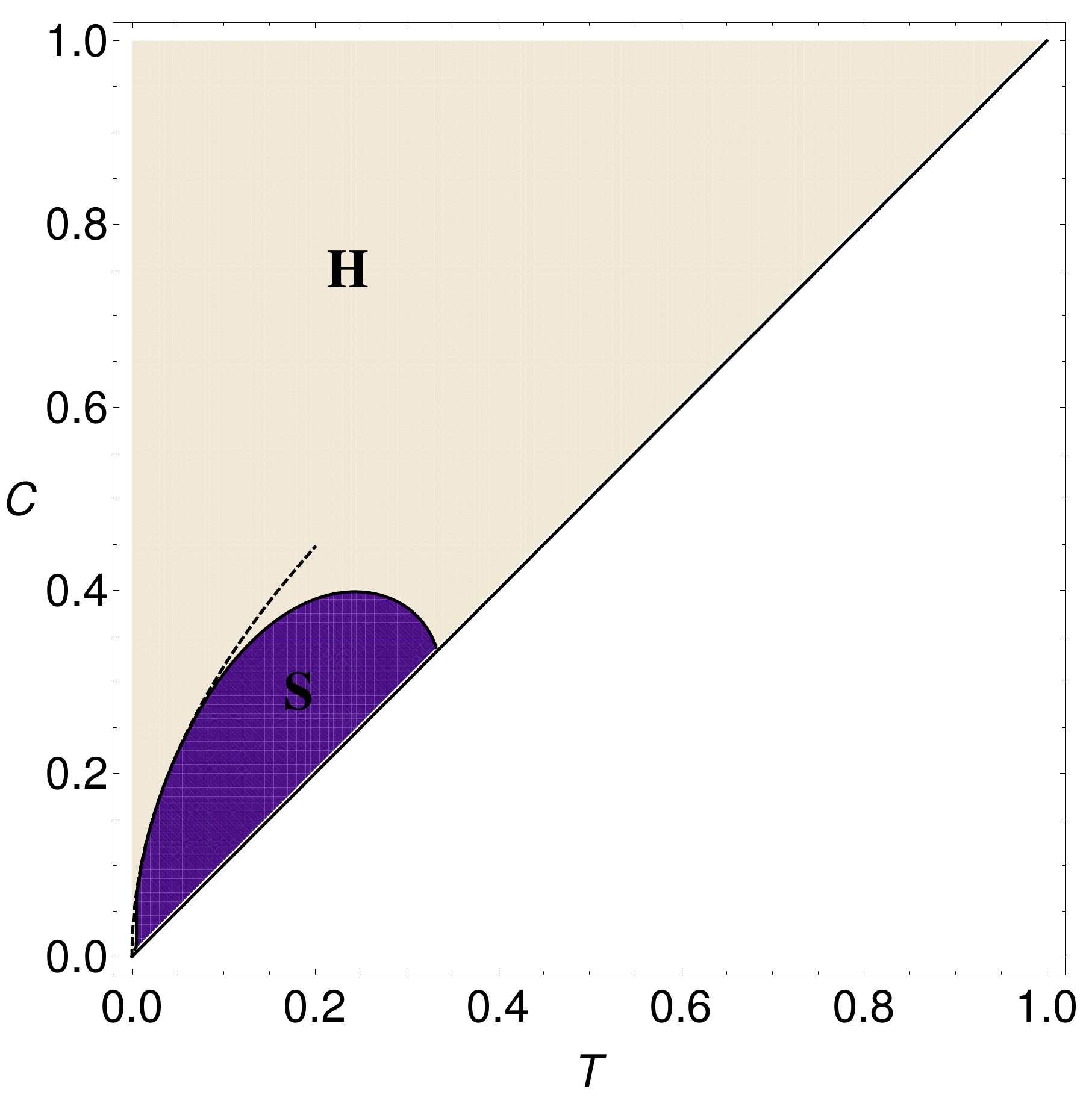}
\caption{Phase diagram in the parameter space $(T,c)$ for the case $\delta=\gamma=0$ ($P=2$). The (S) area represents the region where the symmetric mixtures are stable; the (H) region is 
characterised by hierarchical states, while the paramagnetic state (P) is stable for $T>T_c=c$. The (S) region is obtained as the contour plot of the equation 
$\lambda_2=0$ solved numerically together with equations (\ref{eq:system2}) 
at stationarity. The approach to zero is consistent with $c= \sqrt T$ predicted by the theory (dashed line).}
\label{fig:critline2P}
\end{figure}

A contour plot of the critical line $\lambda_2=0$ in the $T$-$c$ place is shown in fig.  \ref{fig:critline2P}.
We highlight three regions: a paramagnetic region (${\bf P}$) for $T>T_c=c$ where the only solution is 
${\bf m}=0$; a region (${\bf S}$) for $T<c$ and for $c<1/3$ where symmetric solutions are stable; finally a region (${\bf H}$) where $\bm\neq 0$ and symmetric mixtures are unstable, 
so that a new class of asymmetric solutions takes over, the so-called hierarchical solutions.
For $T\to 0$ these take the form $\bm=(1,1-c)$, which is the steady state of the system (\ref{eq:system2}) for $\beta\to\infty$.
Calculating the eigenvalues $\lambda_1,\lambda_2$ of the stability matrix (\ref{eq:Jacobian}) evaluated at $\bm^\star=(1,1-c)$,
we have $\lambda_1,\lambda_2\to -1$ as $\beta\to \infty$, so $\bm=(1,1-c)$ is 
an attractor at low temperature.

\begin{figure}
\begin{picture}(100,620)(5,5)
 \put(10,410){\resizebox{0.4\columnwidth}{!}{\includegraphics{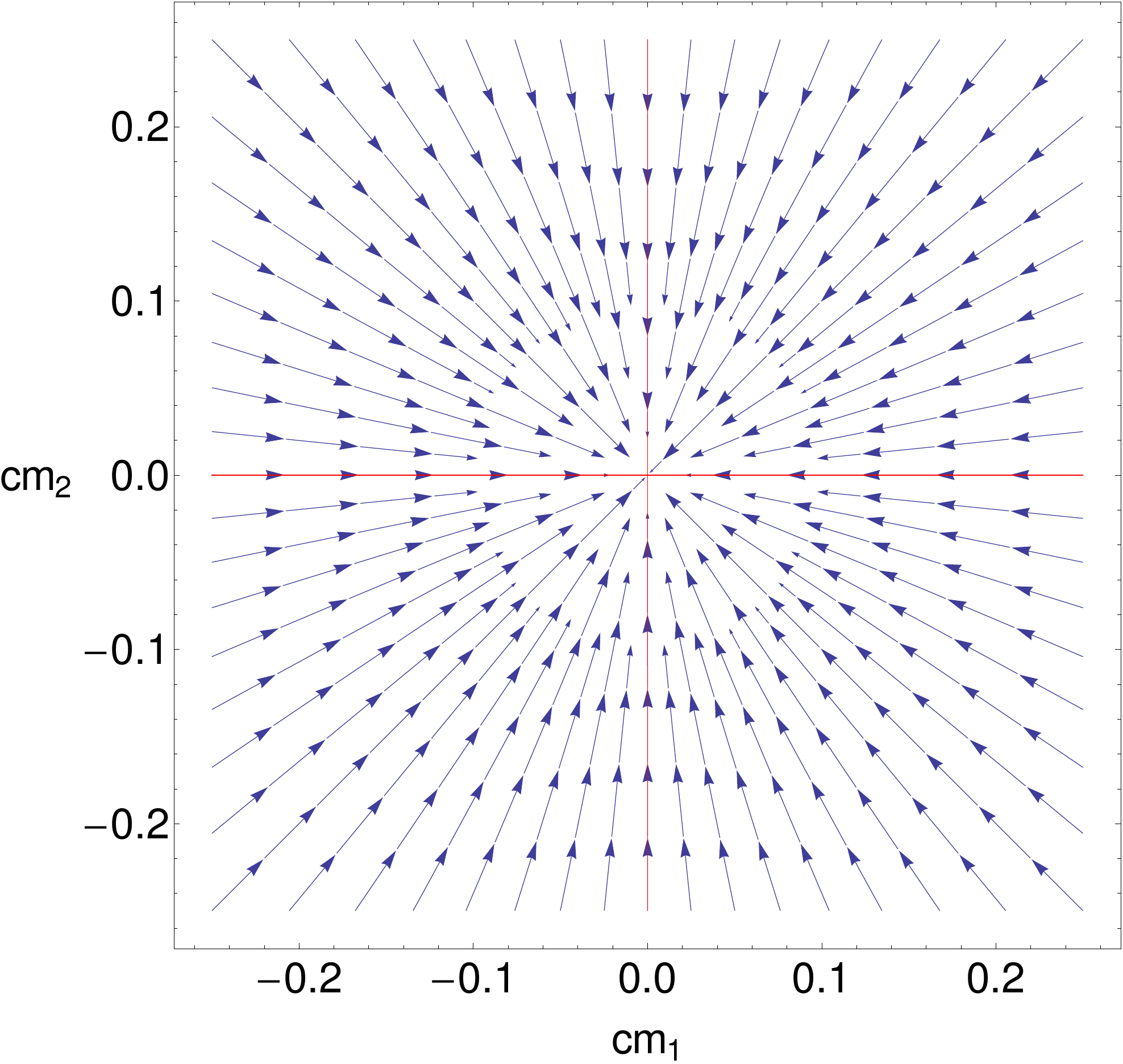}}	
\resizebox{0.58\columnwidth}{!}{\includegraphics{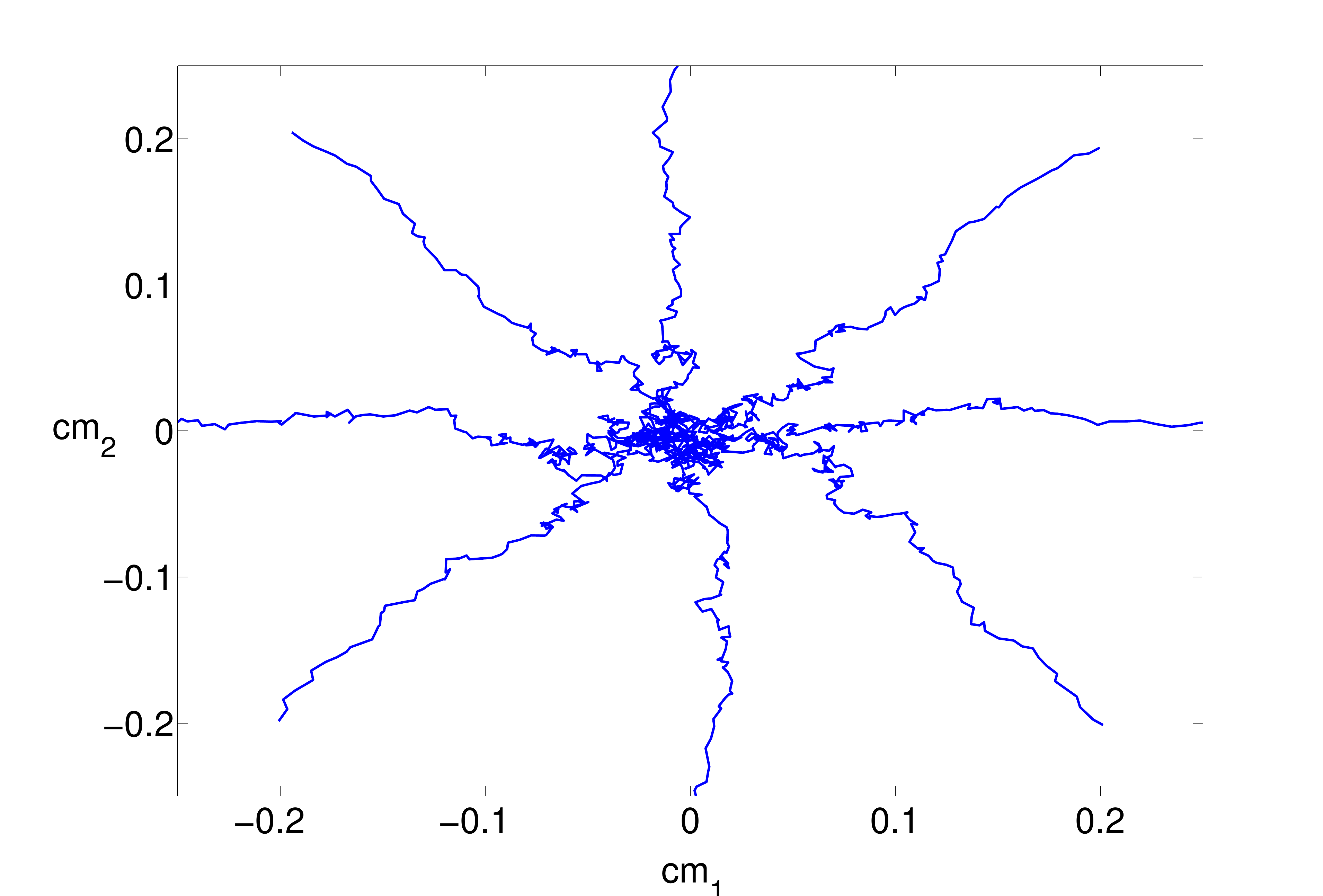}}}
 \put(10,210){\resizebox{0.4\columnwidth}{!}{\includegraphics{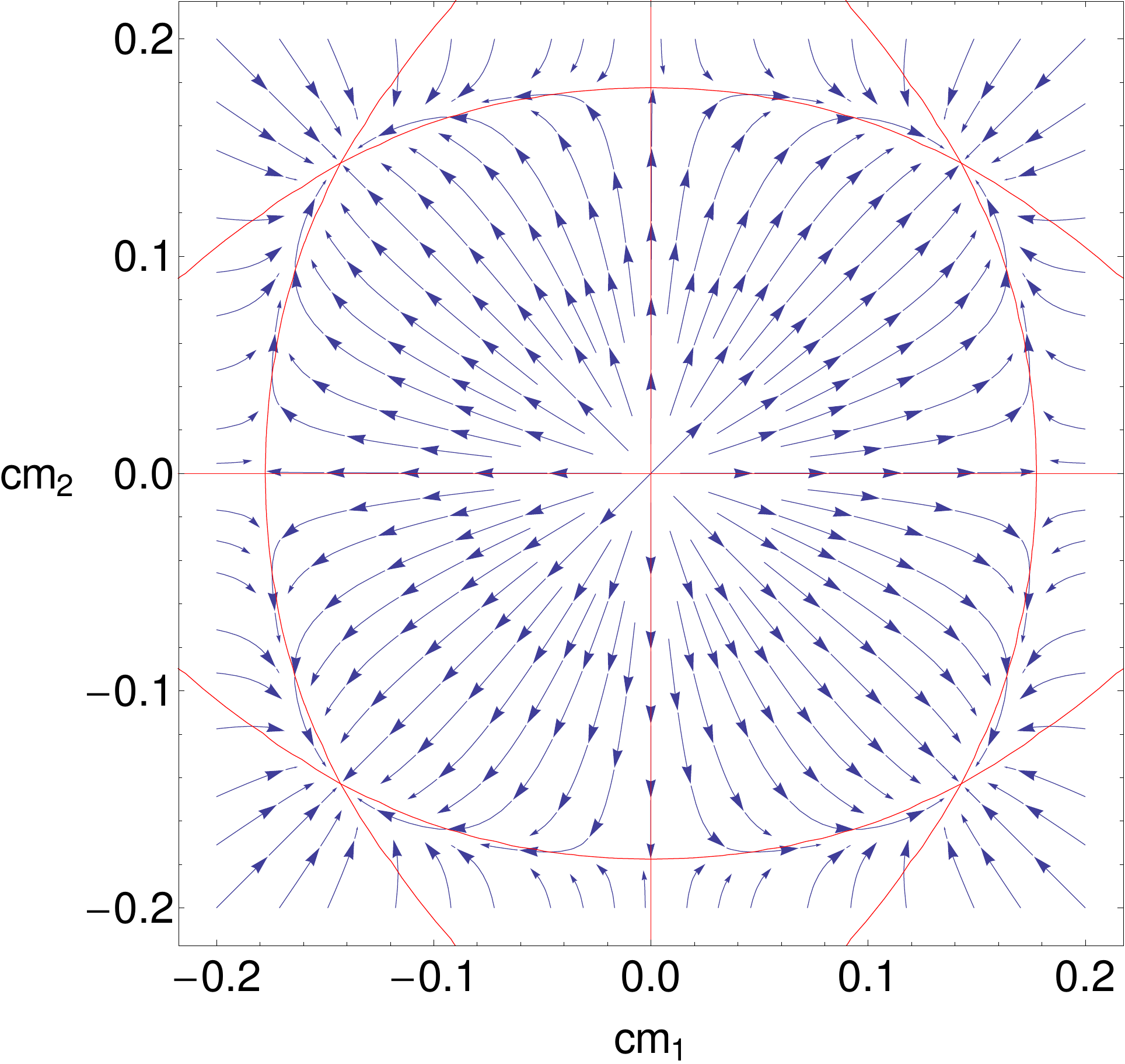}}
\resizebox{0.58\columnwidth}{!}{\includegraphics{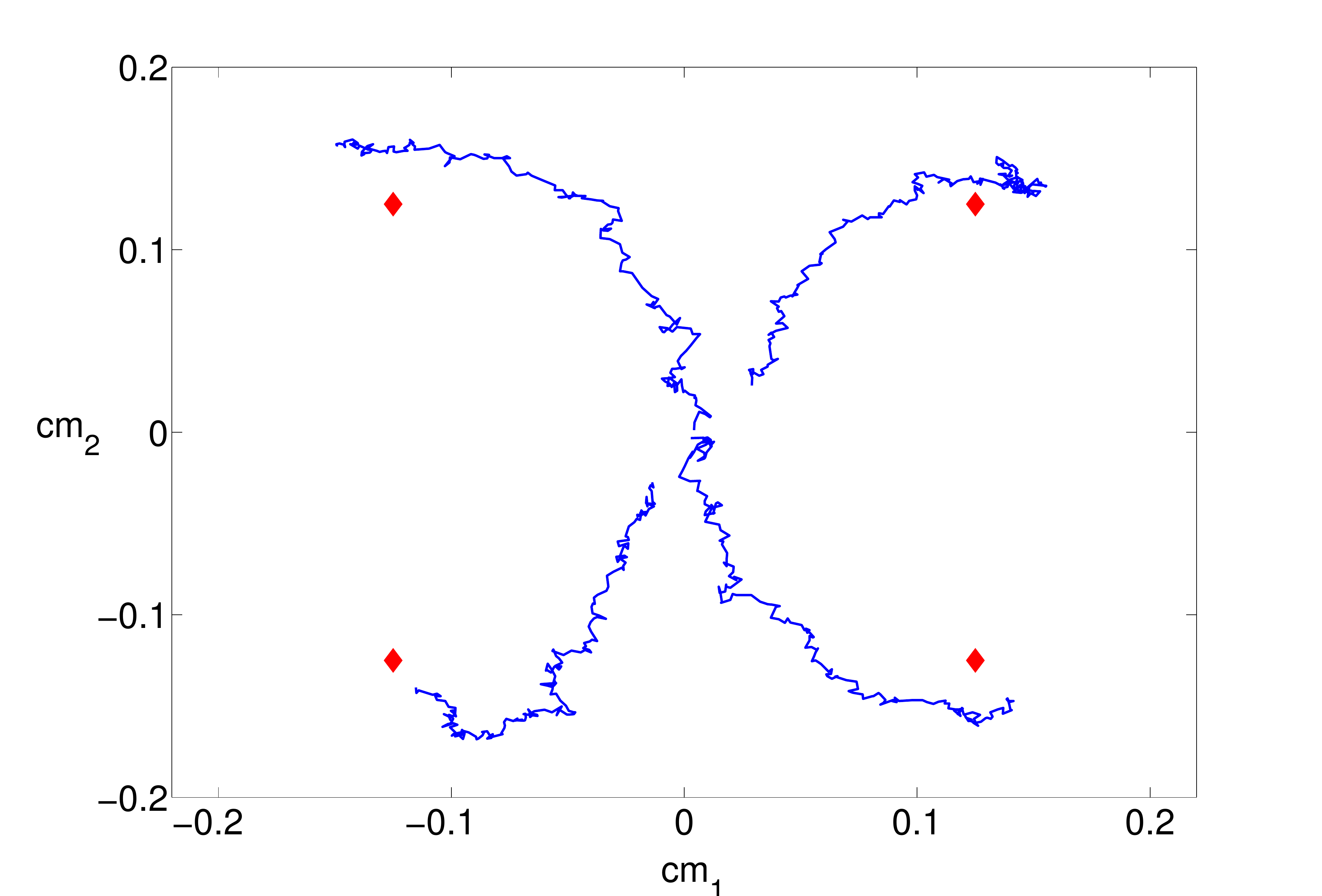}}}
\put(10,10){\resizebox{0.4\columnwidth}{!}{\includegraphics{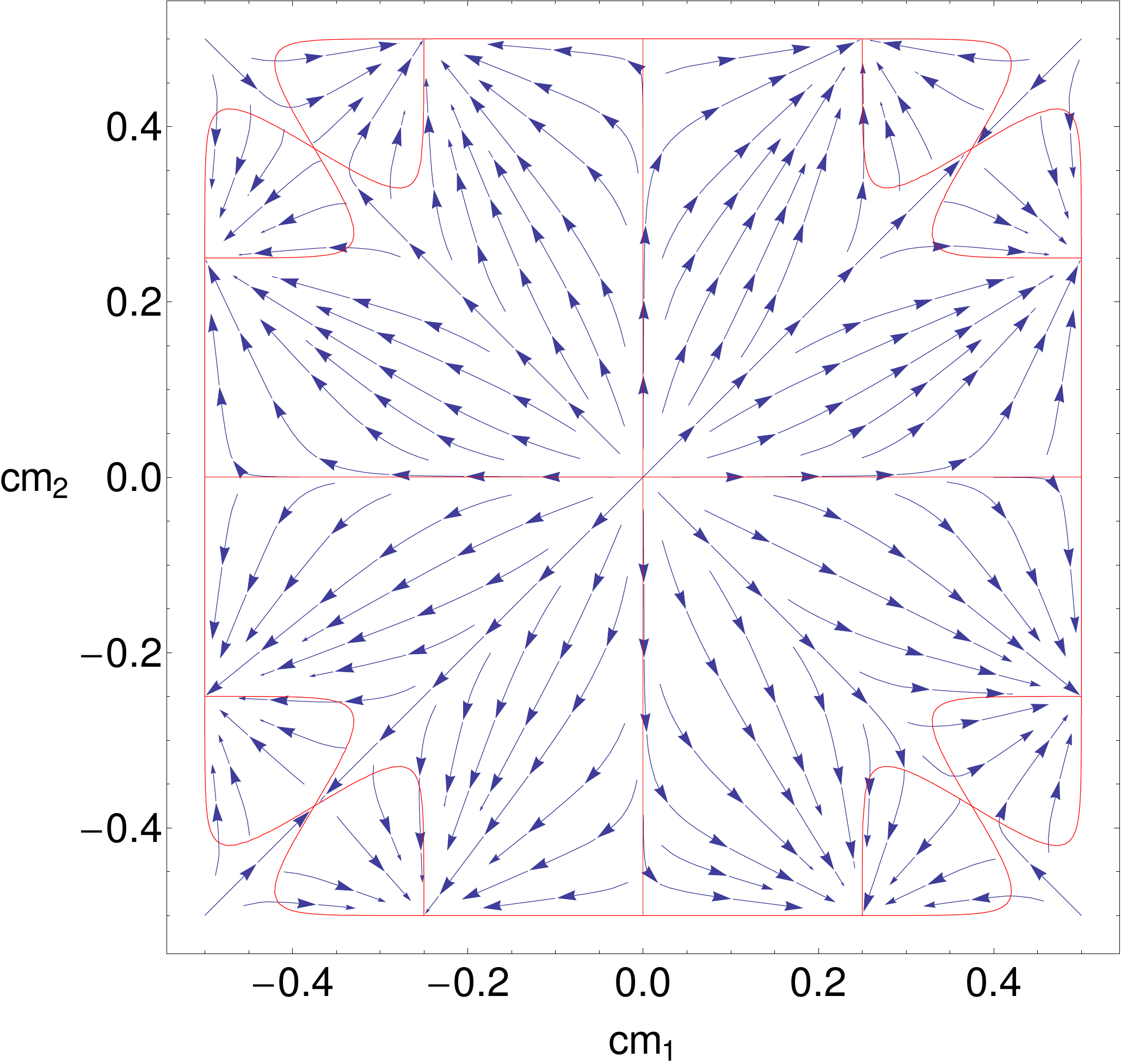}}	
\resizebox{0.58\columnwidth}{!}{\includegraphics{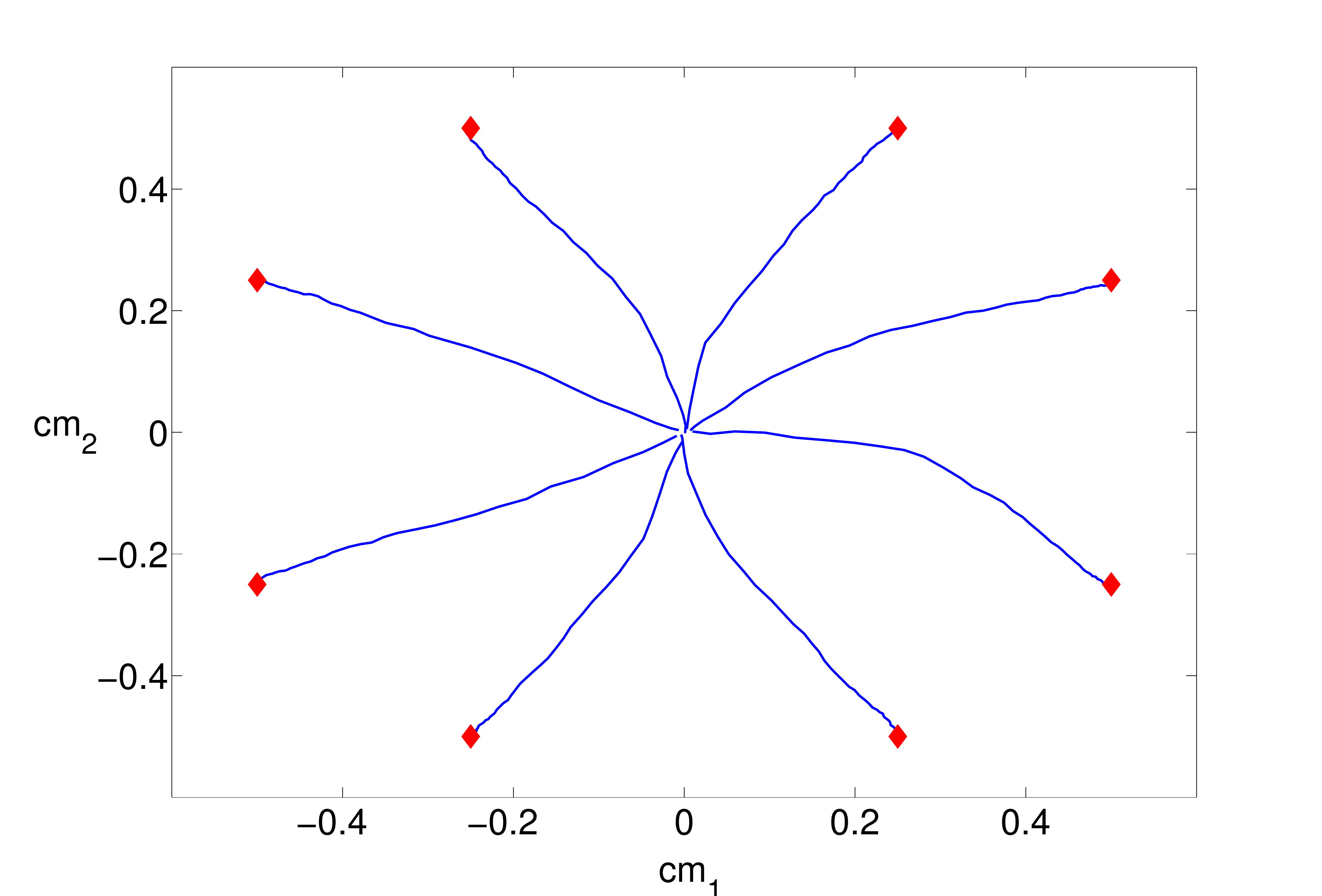}}}
\end{picture}
\hspace*{-5cm}
\caption{Left panels: Phase portraits of the dynamical system (\ref{eq:system2}) in different 
regimes of noise $T$ and dilution $c$; red lines
represent the null-clines, and stationary states are at the intersections of null-clines. Right
panels: Monte-Carlo simulations with $N = 10^4$ spins, with $T$ and $c$ as in the left panels; red markers represent the stationary states of the dynamical system. 
From top to bottom: $c=0.4, \hat T=1.25$; 
$c = 0.25, \hat{T} = 0.8$; $c = 0.5$, $\hat{T} = 0.01$: in the latter regime, 
stationary states are given by $\bm=(1,1-c)$, as expected at low temperature.}
\label{fig:gamma0}
\end{figure}

This behaviour is confirmed by Monte Carlo simulations and by
the phase portrait of the dynamical system \eqref{eq:system2}, both shown 
in fig.  \ref{fig:gamma0}. One can see that phase curves for $\hat T=1.25$, i.e. $T>T_c$, evolve to the steady state $\bm=0$;  
phase curves at temperature $\hat{T} = 0.8$ and $c = 0.25$, (inside the ({\bf S}) region), 
evolves to symmetric mixtures (where $m_1=m_2$); at smaller temperature, $\hat T=0.01$, and $c=0.5$ (inside the ({\bf H}) region), symmetric states 
become unstable with phase velocities pointing away from them, 
and non-symmetric stable steady states (with $m_1\neq m_2$) appear.
At $\hat T=0.01$, the system evolves towards the low temperature hierarchical state ${\bf m}= (1,1-c)$. 
Results are in agreement with equilibrium analysis and simulations 
\cite{hierar, stabilitya, PRE}.

\subsection{Generalisation to $P>2$ patterns}\label{sec:p3}
\begin{figure}
\centering
\resizebox{0.45\columnwidth}{!}{
\includegraphics{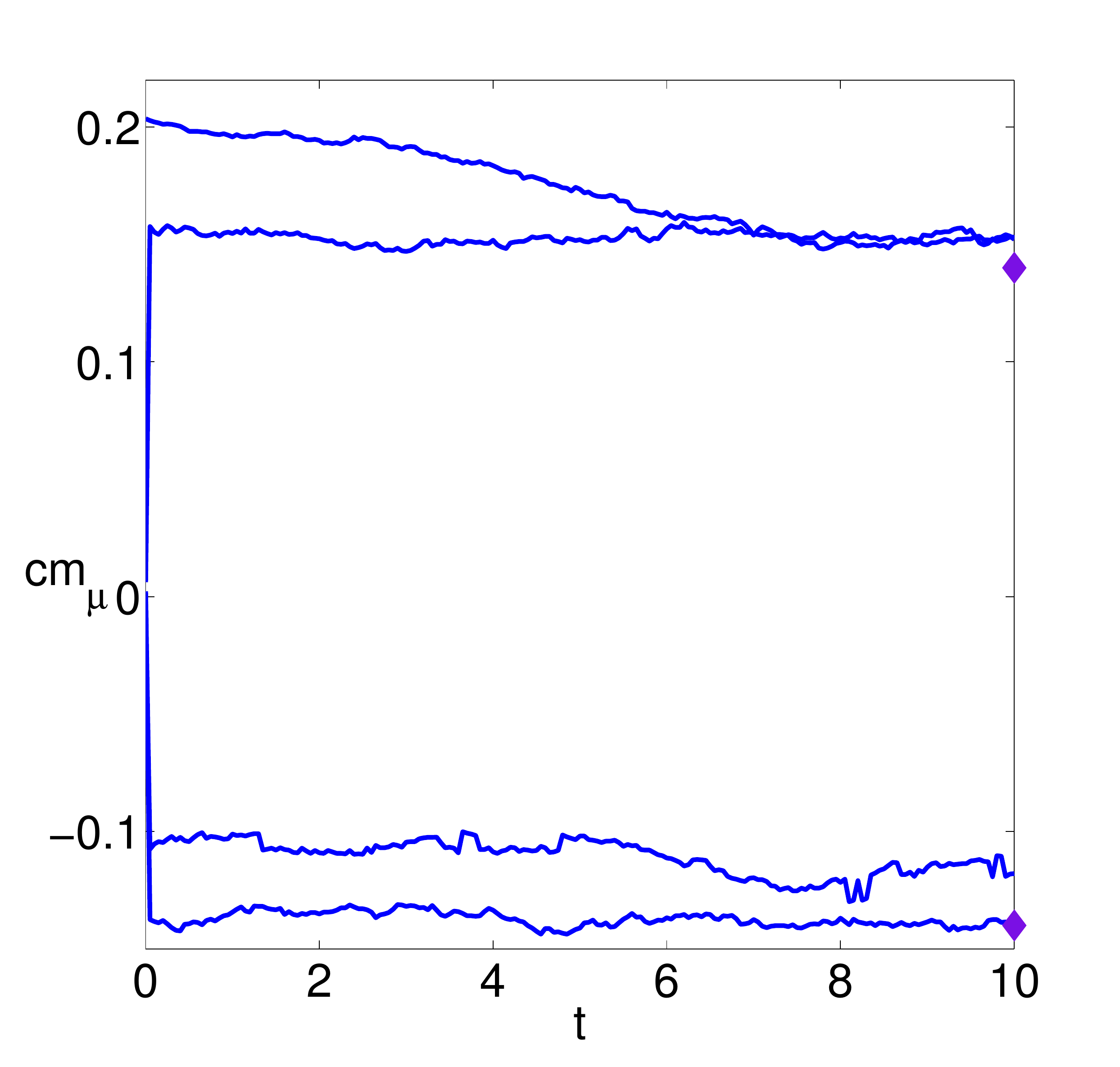}}
\resizebox{0.45\columnwidth}{!}{
\includegraphics{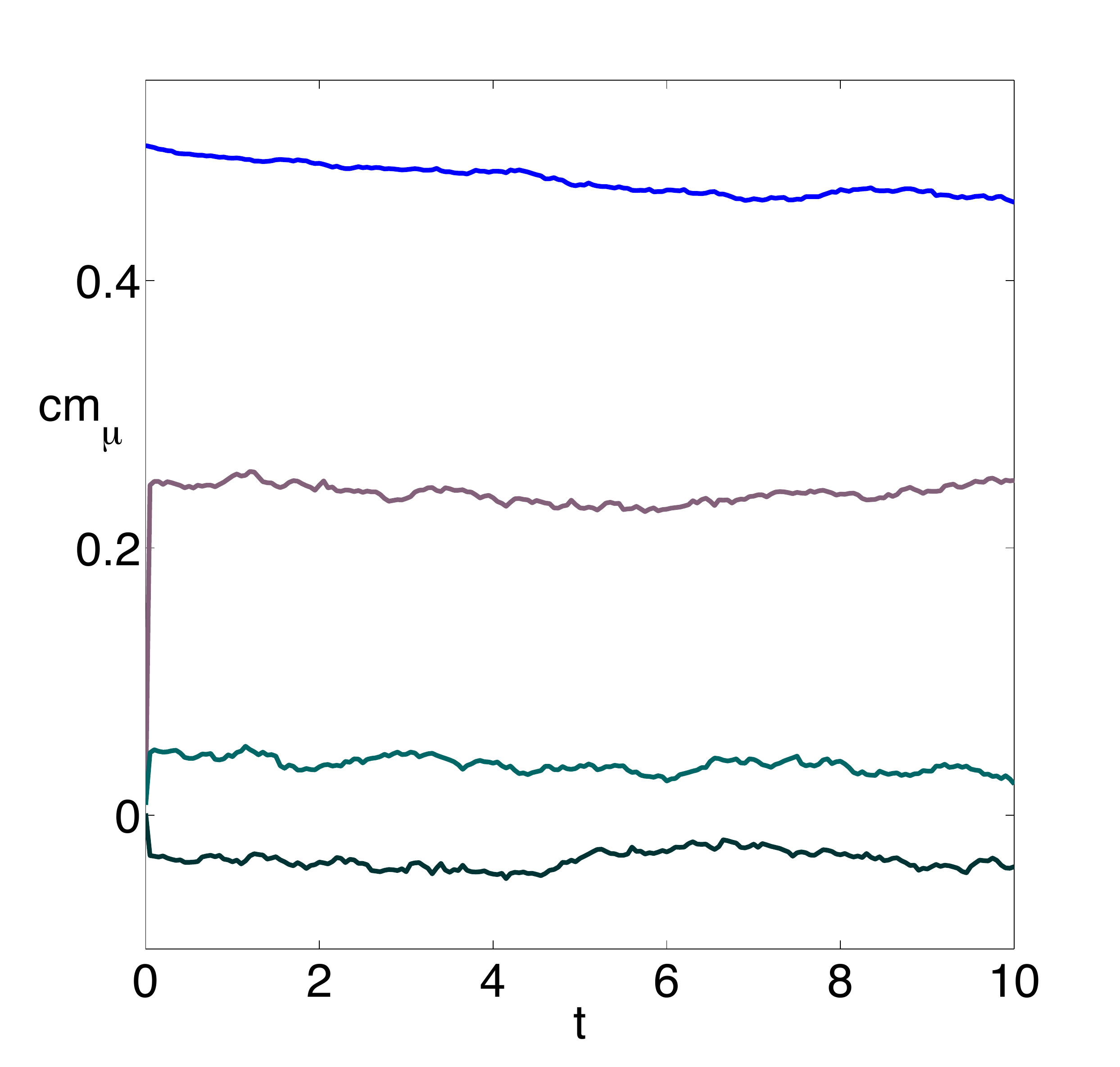}}
\caption{Simulations with $N=10^4$ spins and $P=4$ patterns of  $m_{\mu}$ as a function of time. Left Panel: Symmetric solutions for $\hat{T}=0.5$, $c=0.2$. The markers represent the amplitude of the symmetric mixtures evaluated solving the self-consistency equation for $P=4$ \eqref{eq:selfp4}. Right Panel: hierarchical retrieval for $c=0.5$ and $\hat{T}=0.4.$ }
\label{fig:MC_p4_Tc}
\end{figure}

Generalising to $P$ patterns, and assuming we have, close to criticality, 
$\bm=m(1,\ldots,1,0,\ldots,0)$ with $\bm^2=nm^2$, we get from 
(\ref{eq:mgz})
\begin{eqnarray}
m^2=\frac{3\epsilon}{1-3c+3cn},
\end{eqnarray}
that coincides with \eqref{eq:symmixt} for $P=n=2$. 
To study the stability of symmetric mixtures, we evaluate the 
Jacobian (\ref{eq:Jacobian}) at the steady state $\bm^\star=m(1,\dots,1,0,\dots,0)$. 
Diagonal elements are given by (\ref{eq:Aqm}), (\ref{eq:Aq}) where,
for symmetric states, 
\bea
q_\mu&=&\langle(\tanh^2(\hat{\beta}  m(1+\sum_{\nu\neq \mu}^n \xi^{\nu}))\rangle_{\bxi}
\label{eq:qa}
\\
q&=&\langle(\tanh^2(\hat{\beta} m \sum_{\nu=1}^n \xi^{\nu})\rangle_{\bxi}
\label{eq:qb}
\end{eqnarray}
whereas off-diagonal elements are, for $\gamma=0$, 
\begin{eqnarray}
A_{\mu\nu}&=& -\hat{\beta} Q,\quad\quad {\rm for~}\mu,\nu\leq n
\nonumber\\
A_{\mu\nu}&=&0,\quad\quad\quad\quad {\rm otherwise}
\end{eqnarray}
with
\begin{eqnarray}
Q=\frac{1}{c}\langle( \xi^{\mu}\xi^{\nu}\tanh^2(\hat{\beta}m\sum_{\nu=1}^n \xi^\nu)\rangle_{\bxi}.
\end{eqnarray}
Due to the symmetry of the problem, 
we have three distinct eigenspaces, with $q_\mu=q_1~\forall~\mu$:
\begin{eqnarray}
{\rm Eigenspaces} \quad\quad && {\rm Eigenvalues}\nonumber\\
\bx=(1,\ldots, 1,0,\ldots,0)
\quad\quad && \lambda_1= \hat{\beta}(1-q_1)-1-\hat{\beta}(n-1)Q\nonumber\\
\bx=(x_1,\ldots,x_n,0,\ldots,0),\quad {\mbox {$\sum_\mu x_\mu=0$}} 
\quad\quad && \lambda_2= \hat{\beta}(1-q_1)-1+\hat{\beta}Q\nonumber\\
\bx=(0,\ldots,0,n_{n+1},\ldots,x_P)  \quad\quad &&
\lambda_3= \hat{\beta}(1-q)-1\nonumber
\end{eqnarray}
Near $T\simeq T_c$ we Taylor expand $q_1, q, Q$ in powers of $t=\hat{\beta}-1$, 
obtaining respectively:
\begin{eqnarray}
q_1= \frac{3\epsilon +3 (n-1)c\epsilon}{1-3c+3cn}\\
q= \frac{3 nc\epsilon}{1-3c+3cn}\\
Q= \frac{6\epsilon c}{1-3c+3cn}.
\end{eqnarray}
Hence, eigenvalues near the critical temperature are given by
\begin{eqnarray}
\lambda_1=-\frac{-2\epsilon+6c\epsilon-6c\epsilon n}{1-3c+3cn}\\
\lambda_2= \frac{-2\epsilon+6c\epsilon}{1-3c+3cn}\\
\lambda_3=\frac{\epsilon-3c\epsilon}{1-3c+3cn}.
\end{eqnarray}
We note that $\lambda_1<0$ for any $c$,
and, as in the case $P=2$, $\lambda_2<0$ for $c<1/3$. 
On the other hand,  
$\lambda_3<0$ for $c>1/3$.
Since the eigenvalue $\lambda_3$ only comes into play for $P>n$, 
symmetric mixtures ${\bf m}=m(1,\dots,1)$ ,
where {\it all} patterns are recalled, i.e. with $n=P$, are stable for 
any $c<1/3$, near criticality.
In contrast, there is no region of the phase space 
where mixtures of the form 
${\bf m} =m (1,\dots,1,0,\dots,0)$, in which a number of patterns 
is not recalled, is stable. Results are confirmed by Monte-Carlo 
simulations shown in fig.  \ref{fig:MC_p4_Tc}.
Interestingly, the stability 
threshold $c<1/3$ near $T_c$ is independent of the number of patterns $P$. 
However we note that at other temperatures, the
eigenvalues and their stability will in general depend on the number of 
patterns involved \cite{stabilitya}.

At low temperature, symmetric mixtures become unstable.
In order to show this, it is convenient to introduce the noise distribution 
\bea
P_n(z)&=&\bra \delta(z-\sum_{\nu=1}^{n}\xi^{\nu})\ket_\bxi
=\int_{-\pi}^{\pi} \frac{\rmd\omega}{2\pi} 
\rme^{-\rmi \omega \hat{z}}\prod_{\nu=1}^{n}\langle\rme^{\rmi \omega \xi^{\nu}}\rangle_{\bxi}
\nonumber\\
&=&\int_{-\pi}^{\pi}\frac{\rmd\omega}{2\pi} \rme^{-\rmi\omega z} 
[1+c(\cos\omega -1)]^n
\label{eq:Pn}
\eea
where $n$ is the number of retrieved patterns.
We can express $q_1$ and $q$, as given in (\ref{eq:qm}), (\ref{eq:q}), in terms of 
(\ref{eq:Pn}):
\begin{eqnarray}
q_1&=&\sum_z P_{n-1}(z) \tanh^2(\hat{\beta}m(1+z))
\nonumber\\
q&=& \sum_z P_n(z) \tanh^2(\hat{\beta}m z)
\end{eqnarray}
and the self-consistency equation for the amplitude $m$ of the symmetric 
mixtures $m$ becomes
\begin{eqnarray}
m =\sum_z P_{n-1}(z) \tanh(\hat{\beta}m(1+z)).
\label{eq:selfp4}
\end{eqnarray}
It is easy to see that $P_n(z)$ satisfies a useful recursion relation 
\be
P_n(z)=(1-c)P_{n-1}(z)+c\frac{P_{n-1}(z+1)+P_{n-1}(z-1)}{2}
\label{eq:recursion}
\ee
which represents a discrete-time lazy, symmetric random walker, taking right 
and left 
unit steps along the $z$-axis with equal probabilities $c/2$. 
All the 
$P_n$'s in the hierarchy can then be determined from knowledge of 
$P_0(z)\equiv \delta(z)$. 
The random walker is lazy for $c<1$ because at each iteration $n$ 
it has a finite probability to take no step.
When $\beta \to \infty$, $q_1\to 1- P_{n-1}(-1)$, 
$q\to 1-P_n(0)$ and $Q\to 0$, hence 
\begin{eqnarray}
\lambda_{1,2}\to \hat{\beta}P_{n-1}(1)-1\\
\lambda_3 \to \hat{\beta}P_n(0)-1
\end{eqnarray} 
where we used $P_n(z)=P_n(-z)$.
We note that for $c<1$, we have a non-zero probability 
$P_n(0)>0$ for the walker to be at the origin for any finite time $n\geq 0$, 
hence symmetric mixtures are unstable for 
any value of $n$ at low temperature. In contrast, $c=1$ introduces a 
periodicity in the return times of the walker, 
so that $P_n(0)=0$ for $n$ odd and 
$P_n(1)=0$ for $n$ even, so that for $n$ odd (even) {\it all} eigenvalues are negative (positive),
in agreement with the standard 
(i.e. undiluted) Hopfield model \cite{amit}, 
where, at zero temperature, all odd mixtures are attractors and all 
even mixtures are unstable.  
In particular, for $n=1$, $\lambda_{1,2}=-1$ and 
from (\ref{eq:recursion})
\bea
P_1(z)&=&(1-c)\delta(z)+c\frac{\delta(z+1)+\delta(z-1)}{2}
\eea
so $\lambda_3=\hat\beta (1-c)-1$. Hence, the so-called {\it pure} 
state, where the 
system recalls only one pattern,  
is stable at low temperature
only for the undiluted case $c=1$, but it is unstable for any $c<1$.

In \cite{hierar, stabilitya}, an \emph{ansatz} was made for the form of the 
magnetisation vector in the 
hierarchical state at $T=0$, 
\begin{eqnarray}
m_{\mu}= (1-c)^{\mu-1}\quad\quad \mu=1,\ldots,n
\label{eq:hierarchic}
\end{eqnarray}
where $n\leq P$ is the number of condensed patterns and $m_\mu=0~\forall~\mu>n$.
We shall denote briefly this state as $\bm_H$. We note that (\ref{eq:hierarchic}), with $n=P$, 
are indeed fixed points of (\ref{eq:system2}) for $\beta\to\infty$ and $P=2$, 
and for larger $P$, 
they solve self-consistently (\ref{eq:vecstedy}) for $\beta\to\infty$ 
for wide regions of the dilution parameter $c$.
The stability of this solution is given by the eigenvalues of the stability 
matrix (\ref{eq:Jacobian})
evaluated at $\bm^\star=\bm_H$. The matrix is again diagonal in the limit 
$\beta\to \infty$, with elements given in (\ref{eq:Aqm}), (\ref{eq:Aq})
and
\begin{eqnarray}
q_\mu&=&\langle(\tanh^2(\hat{\beta} (m_\mu+\sum_{\nu\neq \mu}^n \xi^{\nu}m_\nu))\rangle_{\bxi}
\\
q&=&\langle(\tanh^2(\hat{\beta}\sum_{\nu=1}^n \xi^{\nu}m_\mu)\rangle_{\bxi}.
\end{eqnarray}
Upon introducing 
$P_\mu(z)=\bra \delta(z-\sum_{\nu\neq \mu}^n \xi^\nu m_\nu)$ and 
$P(z)=\bra \delta(z-\sum_{\nu=1}^n \xi^\nu m_\nu)$, we can rewrite these as 
\begin{eqnarray}
q_\mu&=&\int \rmd z \tanh^2(\hat{\beta} (m_\mu+z)P_\mu(z)
\\
q&=&\int \rmd z \tanh^2(\hat{\beta}z)P(z)
\end{eqnarray}
For $\beta\to\infty$, we have
\begin{eqnarray}
A_{\mu\mu}= \hat{\beta}P_\mu(-m_\mu)-1,  && \quad {\rm for~} \mu\leq n
\\
A_{\mu\mu}=  \hat{\beta}P(0)-1, &&  \quad {\rm for~} \mu>n
\end{eqnarray}
Working out 
\bea
P_\mu(z)
=\int_{-\infty}^{\infty} \frac{\rmd\omega}{2\pi} 
\rme^{-\rmi \omega z}\prod_{\nu\neq \mu}^{n}\langle\rme^{\rmi \omega \xi^{\nu}m_\nu}\rangle_{\bxi}
=\int_{-\infty}^\infty \frac{\rmd\omega}{2\pi} \rme^{-\rmi\omega z} 
\prod_{\nu\neq \mu}^n[1+c(\cos(\omega m_\nu) -1)]
\eea
and 
\be
P(z)=\int_{-\infty}^\infty \frac{\rmd \omega}{2\pi}\, 
\rme^{-i\omega z}
\prod_{\nu=1}^n
[1+c(\cos(\omega m_\nu)-1)]
\ee
again, we can relate $P(z)$ and $P_\mu(z)$, $\forall~\mu$, as follows
\be
P(z)=(1-c)P_{\mu}(z)+c\frac{P_{\mu}(z+m_\mu)+P_{\mu}(z-m_\mu)}{2}
\label{eq:recursion2}
\ee
Setting $z=0$ in the above, we have
\be
P(0)=(1-c)P_\mu(0)+cP_\mu(m_\mu)
\ee
hence stability along the directions $\mu=n+1,\ldots,P$, which requires 
$P(0)=0$, implies $P_\mu(m_\mu)=P_\mu(0)=0~\forall~\mu$ i.e. 
stability along the directions $\mu=1\ldots,n$, since for $0<c<1$ the two 
terms on the RHS are both non-negative.
However, $P(0)=\bra \delta(\sum_\nu \xi^\nu m_\nu)\ket$ 
is bounded from below by the probability of drawing $\bxi={\bf 0}$, i.e.
$P(0)\geq (1-c)^n> 0$, so hierarchic mixtures can 
only be stable for $P=n$. The condition for their stability is $P_\mu(m_\mu)=0 
~\forall~\mu\leq P$. It is easy to see that this condition is satisfied 
for $P=2$, where (\ref{eq:recursion2}) yields 
$P_1(1)=c[\delta(c)+\delta(2-c)]/2$ 
and $P_2(1-c)=(1-c)\delta(1-c)+c[\delta(c)+\delta(2-c)]/2$, both vanishing 
for any $c\in (0,1)$.

\section{Extreme dilution and medium storage regime}\label{sec:medium}
Next we analyse the dynamical behaviour of the system in the medium load regime (i.e. $\delta>0$), 
where the number of patterns scales sub-linearly in the system size, when patterns are extremely diluted (i.e. $\gamma>0$). 

\subsection{Cross-talk effect for $\delta =\gamma$}\label{sec:deltaeqgamma}
In this section we study the case $\delta=\gamma$.
We have shown in Sec. \ref{sec:dynamicaleq} that the first class of stationary states
to bifurcate away from $\bm=0$ below criticality,
are the symmetric mixtures
${\bf m} = m ( 1,\dots, 1, 0, \dots,0)$, with $\bm^2=nm^2$, 
where $n$ is the number of retrieved components. \\
For $\delta=\gamma>0$, 
we assume that $n$ is a fraction of the total number of patterns 
$P=\alpha N^\gamma$, hence we set $n=\phi N^\gamma$, with 
$0\leq \phi \leq \alpha$. Using $\bm^2=\phi N^\gamma m^2$ in (\ref{eq:mgnz}), 
we have  
\begin{eqnarray}
m^2\simeq \frac{3\epsilon}{1+3c\phi}
\label{eq:amplitude}
\end{eqnarray}
Again, to analyse the stability of symmetric mixtures, 
we look at the eigenvalues of the Jacobian (\ref{eq:Jacobian}) evaluated at $\bm=m(1,\dots,1,0,\dots,0)$.
Its elements are given by (\ref{eq:Aqm}), (\ref{eq:Aq}) and (\ref{eq:Aoff}), where, for symmetric states, $q_\mu=q_1~\forall~\mu$ and
\begin{eqnarray}
q_1=\langle \tanh^2(\hat{\beta} m(1+\sum_{\nu\neq \mu}^{\phi N^{\gamma}} \xi^{\nu})\rangle_{\bxi},
\\
q=\langle \tanh^2(\hat{\beta} m \sum_{\nu\neq \mu}^{\phi N^{\gamma}} \xi^{\nu})\rangle_{\bxi}
\label{eq:q2small}
\end{eqnarray}
As before, it is convenient to consider the distribution $P_{\phi}(z)=\langle \delta(z-\sum_{\mu\neq\nu}^{\phi N^{\gamma}}\xi^{\nu})\rangle_{\bxi}$.
In the thermodynamic limit it can be written as
\bea
P_\phi(z)
=\int_{-\pi}^{\pi} \frac{\rmd\omega}{2\pi} \rme^{-\rmi\omega z+
c\phi(\cos\omega -1)}= \rme^{-\phi c} I_z(\phi c)
\label{eq:Pphi}
\eea
where 
$I_z(\phi c)$ is the modified Bessel function of the first kind \cite{Bessel}.
Clearly, $P_\phi(z)=P_\phi(-z)$, and due to the property of the Bessel functions, 
$$
\frac{dI_n(t)}{dt}=\frac{1}{2}[I_{n-1}(t)+I_{n+1}(t)],
$$
$P_\phi(z)$ evolves according to 
$$
\frac{dP_\phi(z)}{d\phi}=\frac{c}{2}[P_\phi(z-1)+P_{\phi}(z+1)-2P_\phi(z)]
$$
which describes a continuous-time symmetric random walker taking unit
steps along the $z$-axis
with rate $c/2$, as $\phi$ increases.
We can rewrite $q_1$ and $q$ in terms of $P_\phi(z)$, as:
\begin{eqnarray}
q_1&=&\sum_z P_{\phi}(z) \tanh^2(\hat{\beta}m(1+z))
\label{eq:q1}
\\
q&=& \sum_z P_\phi(z) \tanh^2(\hat{\beta}m z)
\label{eq:q2}
\end{eqnarray}
In the thermodynamic limit, $N\to\infty$ , $A_{\mu\nu} \to 0$ for $\mu\neq \nu$ and the stability matrix is diagonal, so
that we have two eigenvalues
\begin{eqnarray}
\lambda_1\simeq \hat{\beta}(1-q_1)-1, \quad{\rm with}\quad 
d(\lambda_1) = \phi N^{\gamma}
\label{eq:eigen1}
\end{eqnarray}
\begin{eqnarray}
\lambda_2\simeq \hat{\beta}(1-q)-1, \quad{\rm with}\quad d(\lambda_2)= N^{\gamma} (\alpha -\phi)
\label{eq:eigen2}
\end{eqnarray}
where $d(\lambda_i)$ represents the degeneracy of eigenvalue $\lambda_i$.
Near $T\simeq T_c=c$ we Taylor expand $q_1$ and $q$  in powers of 
$\epsilon=\hat \beta-1$ 
\begin{eqnarray}
q_1\simeq \frac{3\epsilon(1+c\phi)}{1+3c\phi}, \hspace{2cm}q\simeq \frac{3\epsilon c\phi}{1+3c\phi}
\label{eq:retta}
\end{eqnarray}
and compute the eigenvalues
\begin{eqnarray}
\lambda_1\simeq -\frac{2\epsilon}{1+3c\phi}<0, \hspace{2.5cm}\lambda_2\simeq \frac{\epsilon}{1+3c\phi}>0
\end{eqnarray} 
Near $T\simeq 0$, we have
\begin{eqnarray}
q_1\to 1-P_\phi(-1), \hspace{2cm}\lambda_1 \to \hat{\beta}
\rme^{-\phi c}I_1(\phi c)-1
\label{eq:eigenlimit1}
\end{eqnarray}
\begin{eqnarray}
q\to 1-P_\phi(0), \hspace{2.3cm}\lambda_2 \to \hat{\beta}\rme^{-\phi c}
I_0(\phi c)-1
\label{eq:eigenlimit2}
\end{eqnarray}
where the asymptotic holds for large $\hat\beta$.
The functions $I_0(x), I_1(x)$ are non-negative and $\order{(1)}$ for any finite $x>0$, and become 
small only for $x\to \infty$, as $e^{-x}I_n(x)\sim 1/\sqrt{2\pi x}~\forall~n$ for large $x$. 
However, for $\delta=\gamma$,
$\phi\leq \alpha$, hence $\lambda_1, \lambda_2>0$ and symmetric mixtures are unstable for any finite value of 
$\alpha$, at low temperature.

From the analysis above it follows that $\lambda_2>0$ near $T\simeq 0$ and $T\simeq T_c$. 
Indeed, one can check numerically that $\lambda_2$ stays positive $\forall T<c$, for any value of $\phi$, by
computing it from (\ref{eq:eigen2}) and (\ref{eq:q2}), with $m$ solving the self-consistency equation
\begin{eqnarray}
m=\sum_{z}P_\phi(z)\tanh(\hat{\beta} m(1+z)).
\label{eq:self}
\end{eqnarray}
Since $\lambda_2$ only comes into play for $\alpha>\phi$, 
symmetric mixtures bifurcating from ${\bf m}=0$ must be in the form ${\bf m} = m_{\alpha}(1,\dots, 1)$, i.e. with $\phi=\alpha$, indicating parallel retrieval of all patterns.

\begin{figure}
\centering
\includegraphics[width=0.47\textwidth]{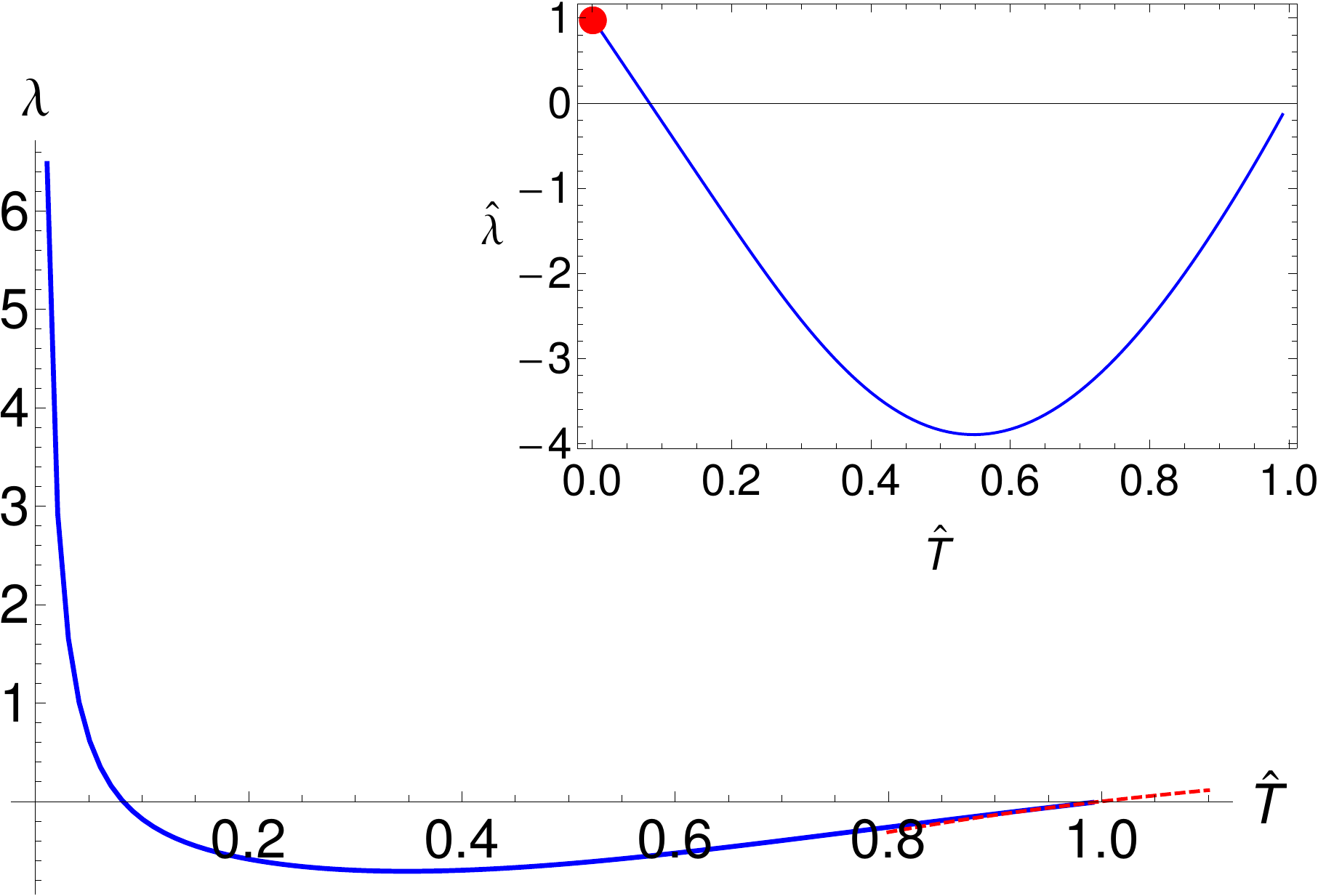}
\caption{Eigenvalue $\lambda_1$ as a function of the temperature $\hat{T}$, for a fixed $\hat{\alpha} = 0.2$. The red dashed line gives the theoretical prediction near the critical temperature (eq.(\ref{eq:retta})). The figure in the inset is in agreement with (eq.(\ref{eq:eigenlimit1})), which holds for $T\to 0$ and gives $\hat{\lambda}=\lambda/\rme^{-\phi c}
I_0(\phi c)\to1$.
}
\label{fig:lambda}
\end{figure}

\begin{figure}
\centering
\includegraphics[width=0.45\textwidth]{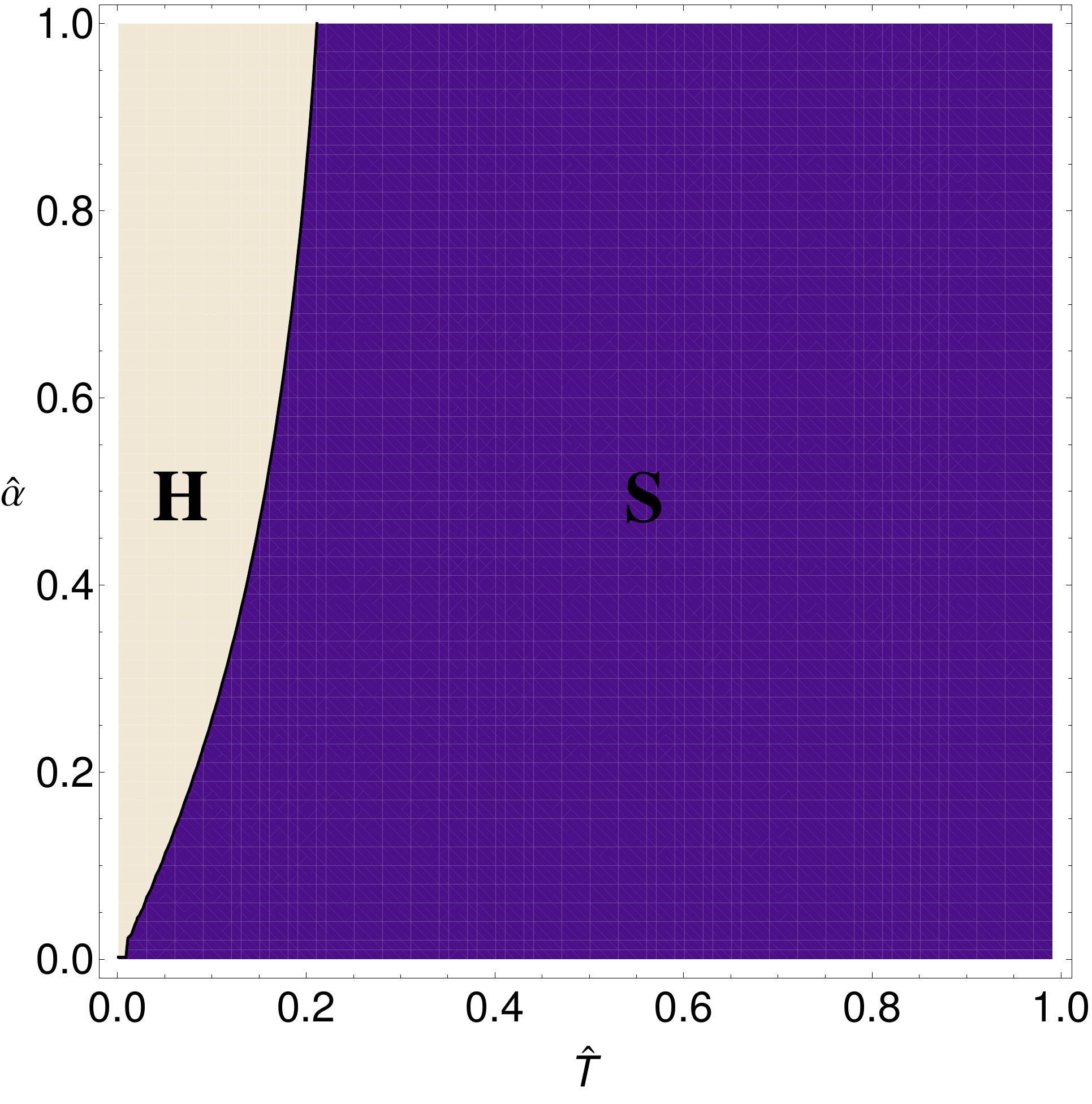}
\caption{Phase diagram in the regime $\delta=\gamma$. In the parameter space $(\hat{T},\hat{\alpha})$ the purple area represents the region where symmetric mixtures exist and are stable, which correspond to $\lambda_1 <0$. 
The area denoted by $H$ represents the region where we expect a hierarchical retrieval.}
\label{fig:phasediagram}
\end{figure}

Their stability is controlled by $\lambda_1$. In fig.  \ref{fig:lambda} we plot the eigenvalue $\lambda_1$ as a function of the scaled temperature $\hat{T}=T/c$ for a fixed value $\hat{\phi}=\alpha c$, 
as well as the theoretical predictions near $T_c$ (\ref{eq:retta}) and $T\simeq 0$ (\ref{eq:eigenlimit1}).  
In fig.  \ref{fig:phasediagram} we show the critical line in the parameter space $\hat{T},\hat{\alpha}$ separating the region ({\bf S}), 
where symmetric mixtures are stable, from the region ({\bf H}) where they are unstable.
The phase diagram shows that for huge regions of the tunable parameters (noise and storage load) 
the system is able to recall all the stored patterns in parallel and symmetrically. However, for low temperature, symmetry of patterns is broken and should not 
be assumed in analysis at zero temperature \cite{medium}. We propose in the next subsection the hierarchical structure of the steady state at zero temperature. 
In fig.  \ref{fig:hierc1} (left panel) we show Monte-Carlo simulations of a system with $N=10^4$ spins, evolving according to sequential Glauber dynamics with $\delta=\gamma=0.25$, $T=0.75$, $c=0.8$ and $\alpha=1$. Such a choice of the parameters corresponds to the region in the phase space where symmetric mixtures are stable. Overlaps with different patterns are seen to evolve to symmetric, non-zero values in agreement, up to finite-size effects, with the values predicted by the theory (\ref{eq:self}). 

In conclusion, pattern cross-talk manifests in this regime as a noise with 
the random-walk distribution (\ref{eq:Pphi}), 
which has the effect of shrinking the amplitude 
of symmetric mixtures with respect to the regime without cross-talk, 
where $P_\phi(z)$ crosses over to a delta function centered in $z=0$ and 
amplitudes are given by the Curie-Weiss equation \eqref{eq:decouple}.

\subsubsection{Extending the hierarchical ansatz to the medium storage 
regime}

\label{sec:lowt2}
As explained above, 
symmetric mixtures are unstable at low temperature.
It is not {\em a priori} clear how the symmetry is broken, 
however for $T\to 0$ one may expect 
that the systems retrieves patterns in a hierarchical fashion similar to 
the one found in the low storage regime.
Reasoning as in \cite{hierar,stabilitya,PRE}, 
we can assume that the system starts
aligning its entries to the non-zero entries of the first pattern, which are $c N^{1-\gamma}$, sparing $N - c  N^{1-\gamma}$ entries of 
$\bsigma$ to align with the $cN^{1-\gamma}$ non-zero entries of the second pattern and so on. Proceeding iteratively, we get to a heuristic general expression
\begin{eqnarray}
m_{\mu}= \bigg (1- \frac{c}{N^{\gamma}}\bigg)^{\mu-1}
\quad\quad \mu=1,\ldots,\phi N^\gamma,
\label{eq:hiergamma}
\end{eqnarray}
\begin{figure}
\centering
\resizebox{0.46\columnwidth}{!}{\includegraphics[width=0.5\textwidth]{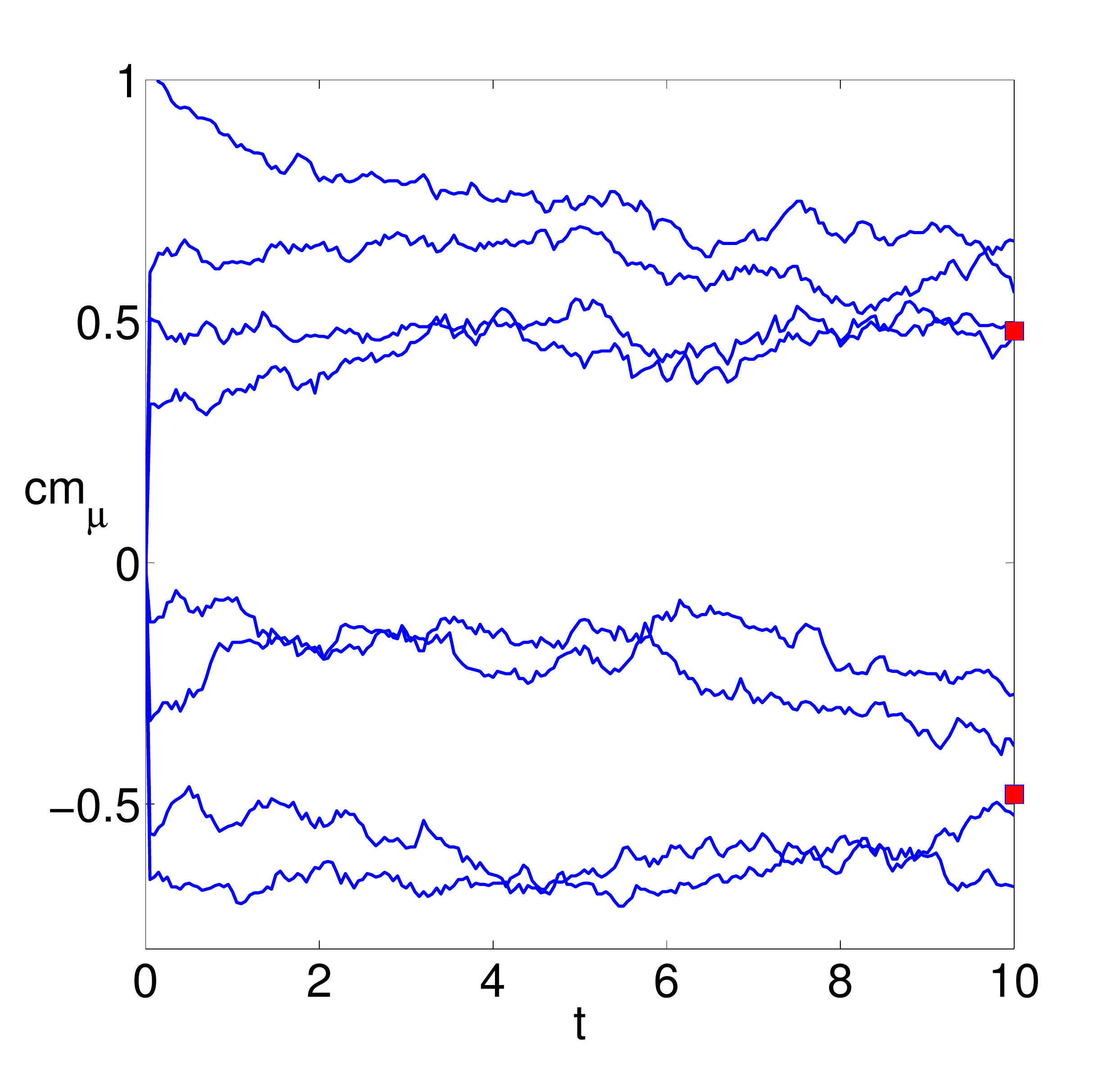}}
\resizebox{0.44\columnwidth}{!}{\includegraphics{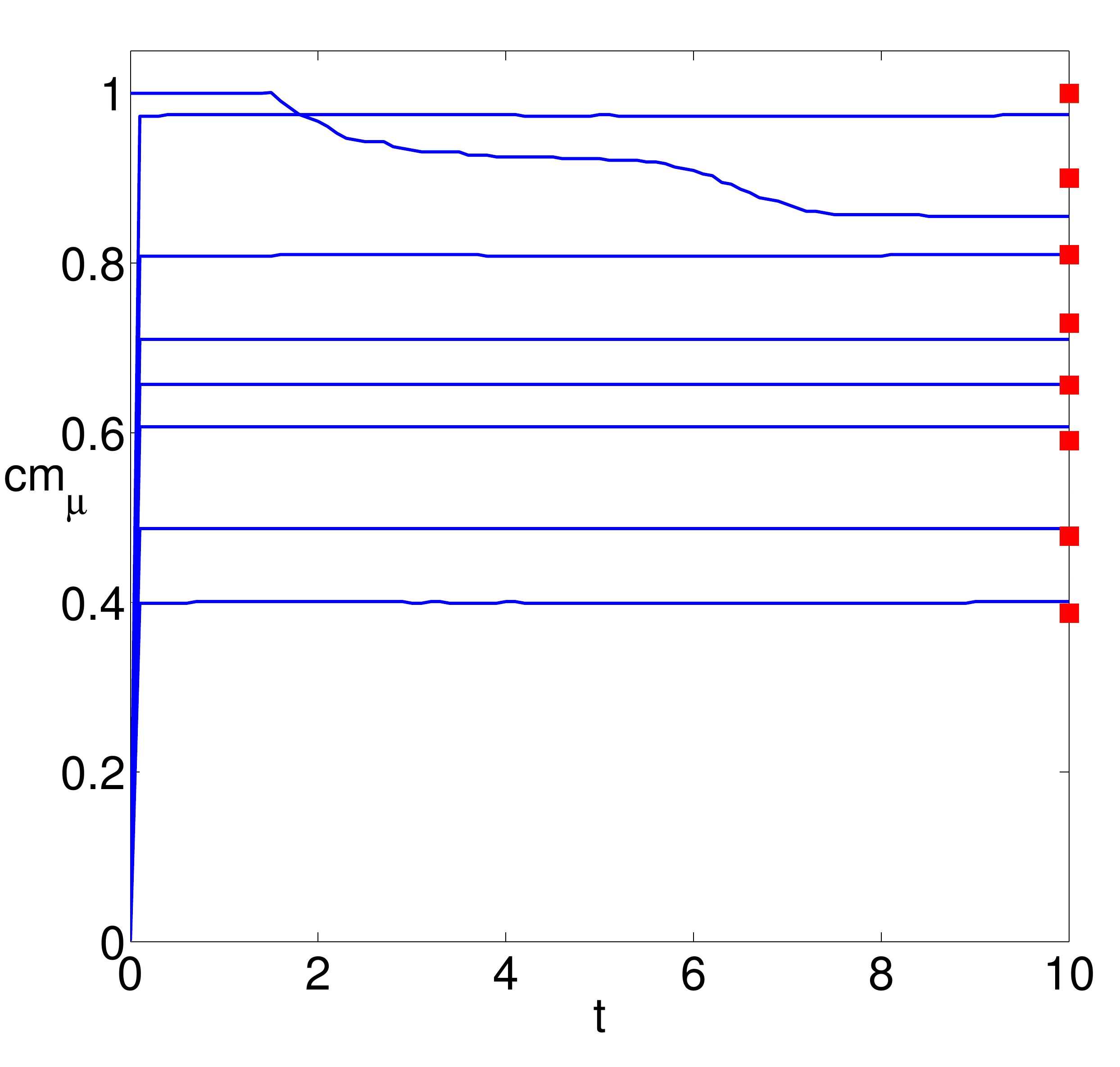}}
\caption{Monte-Carlo simulations of a system with $N=10^4$ spins, $\gamma =\delta =0.25$ and $\alpha=1$ 
evolving according to sequential Glauber dynamics.  We plot the overlaps $m_{\mu}$ with different patterns $\bxi^{\mu}$ as a function of time $t$.
Left: $\hat{T}=0.75$, $c=0.8$. Red markers represent the values theoretically predicted by (\ref{eq:self}). Right: $\hat{T}=0.01$, $c=1$. Red markers 
represent the values heuristically predicted by the ansatz (\ref{eq:hiergamma}) at $T=0$.}
\label{fig:hierc1}
\end{figure}
whose stability at $T=0$ is given by the eigenvalues 
\bea
\lambda_\mu&=&\hat{\beta} P(m_\mu)\quad\quad\mu=1,\ldots,\phi N^\gamma
\nonumber\\
\lambda_\nu&=&\hat{\beta} P(0)\quad\quad\nu=\phi N^\gamma+1,\ldots,\alpha N^\gamma
\eea
where $P$ is given in (\ref{eq:Pgammageq0}).
As in Sec. (\ref{sec:p3}), the probability $P(0)$ is bounded by the 
probability of drawing $ \bxi=0$, $P_{\phi}(0)\geq 
(1-cN^{-\gamma})^{\phi N^\gamma}\simeq \rme^{-c\phi}>0$, hence a necessary condition 
to have stability is $\alpha=\phi$. 
Monte-Carlo simulations at small temperature support the validity of the 
ansatz (\ref{eq:hiergamma}) for $\gamma>0$, 
with $\phi=\alpha$ (see fig. \ref{fig:hierc1}, right panel). 
Remarkably, the first pattern of the hierarchy is recalled without errors, besides the presence of cross-talk.

\subsection{Strong interference for $\delta >\gamma$}\label{sec:deltamaggamma}
In Sec \ref{sec:deltaeqgamma} we have shown that solutions in the form 
$\bm=m(1,\ldots,1,0,\ldots,0)$ were always 
unstable unless $\phi N^\gamma=P$. Hence, for $\delta>\gamma$ we assume that 
the number of condensed patterns is 
$n=\psi N^\delta$.
The amplitude of symmetric mixtures follows from (\ref{eq:self}) as
\begin{eqnarray}
m= \sum_{z=-\alpha N^\delta}^{\alpha N^\delta} P_\psi(z)
\tanh[\hat{\beta}m(1+z)]
\label{eq:self_psi}
\end{eqnarray}
where 
\bea
P_\psi(z)=
\int_{-\pi}^{\pi} \frac{\rmd\omega}{2\pi} \rme^{-\rmi\omega z+
c\psi N^{\delta-\gamma}(\cos\omega -1)}.
\label{eq:ppsi}
\eea
For large $N$, the integral above has most of its mass concentrated around $\omega=0$, 
hence we can use the small-$\omega$ expansion in the exponent
\bea
P_\psi(z)&=&\int_{-\pi}^{\pi} \frac{\rmd\omega}{2\pi} \rme^{-\rmi\omega z-
\frac{c\psi N^{\delta-\gamma}}{2}\omega^2}=
\frac{\rme^{-z^2/2c\psi N^{\delta-\gamma}}}{\sqrt{2\pi c \psi 
N^{\delta-\gamma}}}
\label{eq:Ppsi}
\eea
where in the second equality we rescaled $\omega \to \omega/ \sqrt{c\psi N^{\delta-\gamma}}$ and 
extended the rescaled boundaries to infinity, due to the fast decay 
of the exponential at large values of $\omega$, and performed the Gaussian 
integral.
Hence, the cross-talk between the large number of insufficiently diluted 
condensed patterns manifests as a Gaussian noise, in this regime, 
as expected from the asymptotics 
of the random-walk distribution for large times. 
The variance of the Gaussian noise is given by $\hat \phi=c\phi$ where 
$\phi=\psi N^{\delta-\gamma}$ quantifies the level of interference between 
patterns.
The effect of interference on the symmetric mixtures is stronger in this regime, leading to smaller 
amplitudes of the symmetric recall. 
We anticipate, however, that in the limit of large interference, 
$c\phi\gg 1$, symmetric mixtures are unstable and 
a parallel recall is still accomplished by the system, in contrast with predictions in \cite{medium}, 
where symmetric mixtures were assumed. 

\begin{figure}
\centering
	\includegraphics[width=0.5\textwidth]{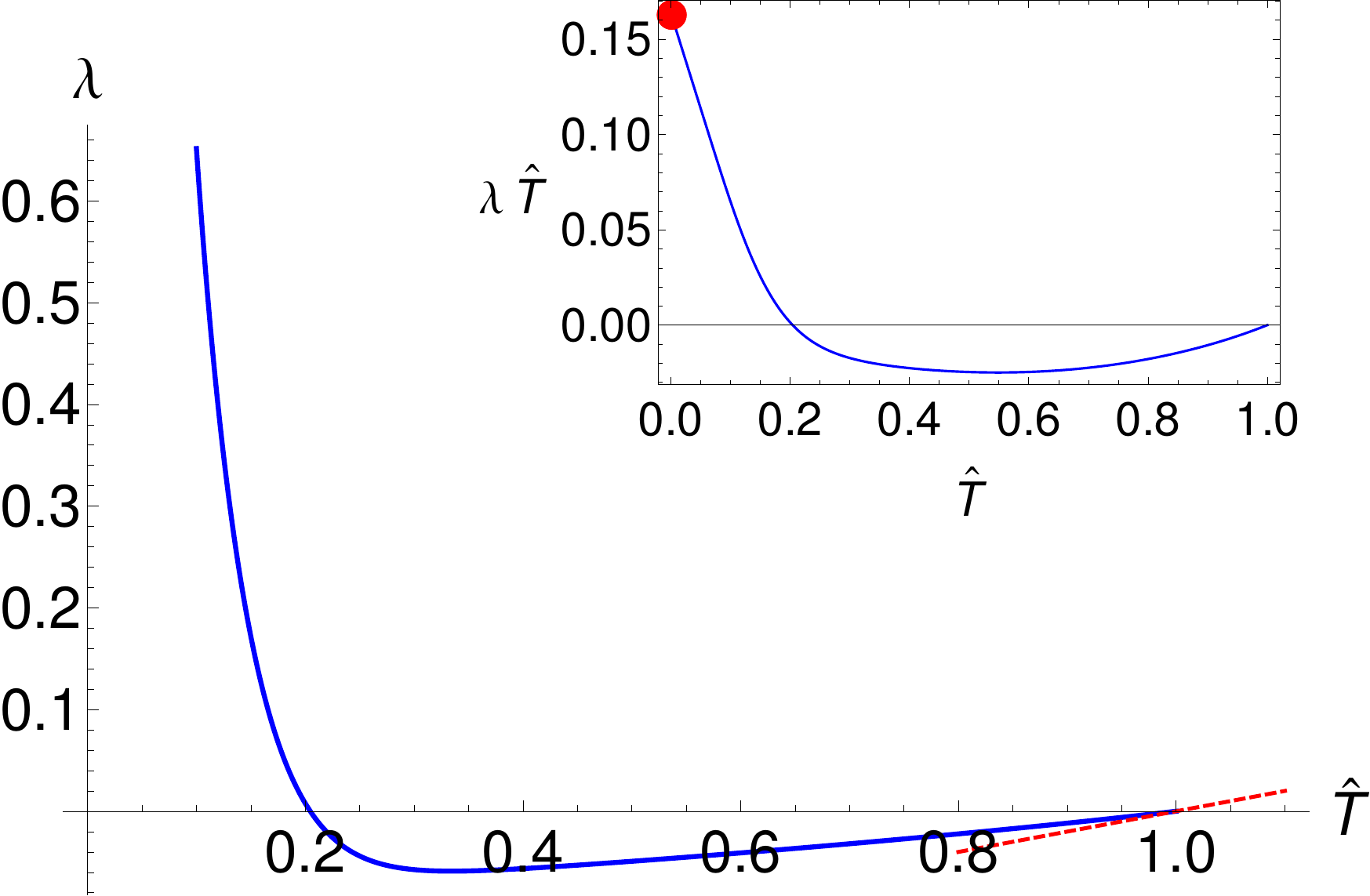}
\caption{Eigenvalue $\lambda_1$ as a function of the temperature $\hat{T}$, for a fixed $\hat{\phi}=c\phi =5$. The red dashed line gives the theoretical prediction near 
the critical temperature (\ref{eq:l1}). The figure in the inset shows agreement with (\ref{eq:l3}), which 
gives $\lambda_1/(\hat T) \to \frac{
N^{(\gamma-\delta)/2}}{\sqrt{2\pi c \psi}}$ as $\hat{T}\to 0$ (red marker).}
\label{fig:eigensecbig}
\end{figure}

Linear stability analysis of the symmetric mixtures $\bm=m(1,\ldots,1,0\ldots,0)$ 
with $n=\psi N^\delta$ can be performed as in Sec \ref{sec:deltaeqgamma}, by identifying $\phi=\psi N^{\delta-\gamma}$.
For $T\simeq T_c$, we get from equations 
(\ref{eq:eigen1}) and (\ref{eq:eigen2}), 
\bea
q_1\simeq \epsilon\Big(1+\frac{2}{3c\psi N^{\delta-\gamma}}\Big),\quad\quad\quad
q\simeq \epsilon\Big(1-\frac{1}{3c\psi N^{\delta-\gamma}}\Big)
\eea
which give
\bea
\lambda_1\simeq -\epsilon\frac{1}{c\psi N^{\delta-\gamma}},\quad
\label{eq:l1}
\quad\quad \lambda_2 \simeq \epsilon\frac{1}{3c\psi N^{\delta-\gamma}}.
\label{eq:l2}
\eea 
For any large but finite $N$, symmetric mixtures are stable 
only if all components are recalled, as 
$\lambda_1<0$ and $\lambda_2>0$. In the thermodynamic limit, (\ref{eq:l2}) vanish, and stability cannot be assessed to linear order in $\epsilon$.
In the low temperature limit, $T\to 0$, we have, similarly to (\ref{eq:eigenlimit1}),
(\ref{eq:eigenlimit2}),
\begin{eqnarray}
q_1\to 1-P_\psi(-1), \hspace{2cm}\lambda_1 \to \hat{\beta}P_\psi(1)
\label{eq:t0big}
\end{eqnarray}
\begin{eqnarray}
q\to 1-P_\psi(0), \hspace{2.3cm}\lambda_2 \to \hat{\beta}P_\psi(0).
\end{eqnarray}
For large $N$, $P_\psi(1)\simeq P_\psi(0) 
\simeq N^{(\gamma-\delta)/2}/\sqrt{2\pi c \psi}$ and we get
\be
\lambda_1\simeq \lambda_2 \to \frac{\hat{\beta} 
N^{(\gamma-\delta)/2}}{\sqrt{2\pi c \psi}}
\label{eq:l3}
\ee
hence for any large and finite $N$, both eigenvalues are positive and symmetric mixtures are unstable, indicating that a hierarchical retrieval must take place.
We can compute $\lambda_{1,2}$ numerically from (\ref{eq:eigen1},
\ref{eq:eigen2}) 
by using (\ref{eq:q1},\ref{eq:q2}) 
where $P_\phi$ is replaced with (\ref{eq:Ppsi}) and the amplitude $m$ 
is determined from (\ref{eq:self_psi}).
In Fig.~\ref{fig:eigensecbig} we plot the eigenvalue $\lambda_1$ as a function of 
$\hat T$ at a large fixed value of $\hat \phi$, as well as the predicted limits for $\hat{T}\to 0$ 
and $\hat{T}\to 1$ given in \eqref{eq:l1} and \eqref{eq:l3} respectively. We check numerically that $\lambda_2$ stays positive for 
any temperature below criticality.

\begin{figure}
\centering
	\includegraphics[width=0.45\textwidth]{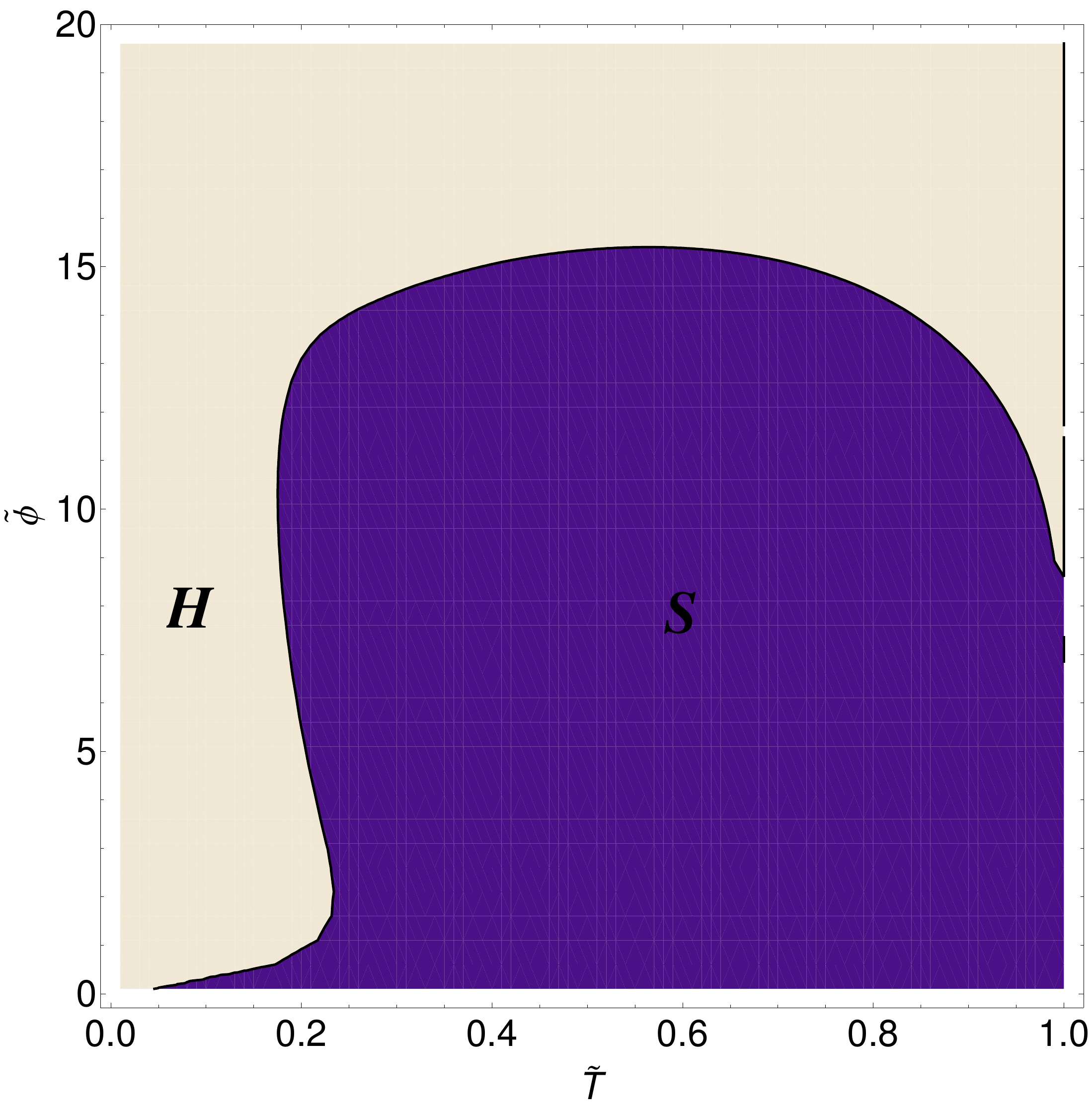}
\caption{Phase diagram in the parameter space $(T,c)$ for the case $\delta>\gamma$. The (S) area represents the region where the symmetric mixtures are stable, while the (H) region is 
characterised by hierarchical states. The (S) region is obtained as the contour plot of the equation 
$\lambda_1=0$ solved numerically together with equation (\ref{eq:self_psi}).}
\label{fig:phibig}
\end{figure}

A contour plot of the critical line where $\lambda_1=0$ in the $T-c$ place is shown in Fig. \ref{fig:phibig}. Again, we identify the region S where symmetric mixtures are stable and 
an H region where these become unstable. We note that for strong interference or low temperature the symmetry of pattern is broken.
\begin{figure}
\begin{center}
\resizebox{0.45\columnwidth}{!}{\includegraphics{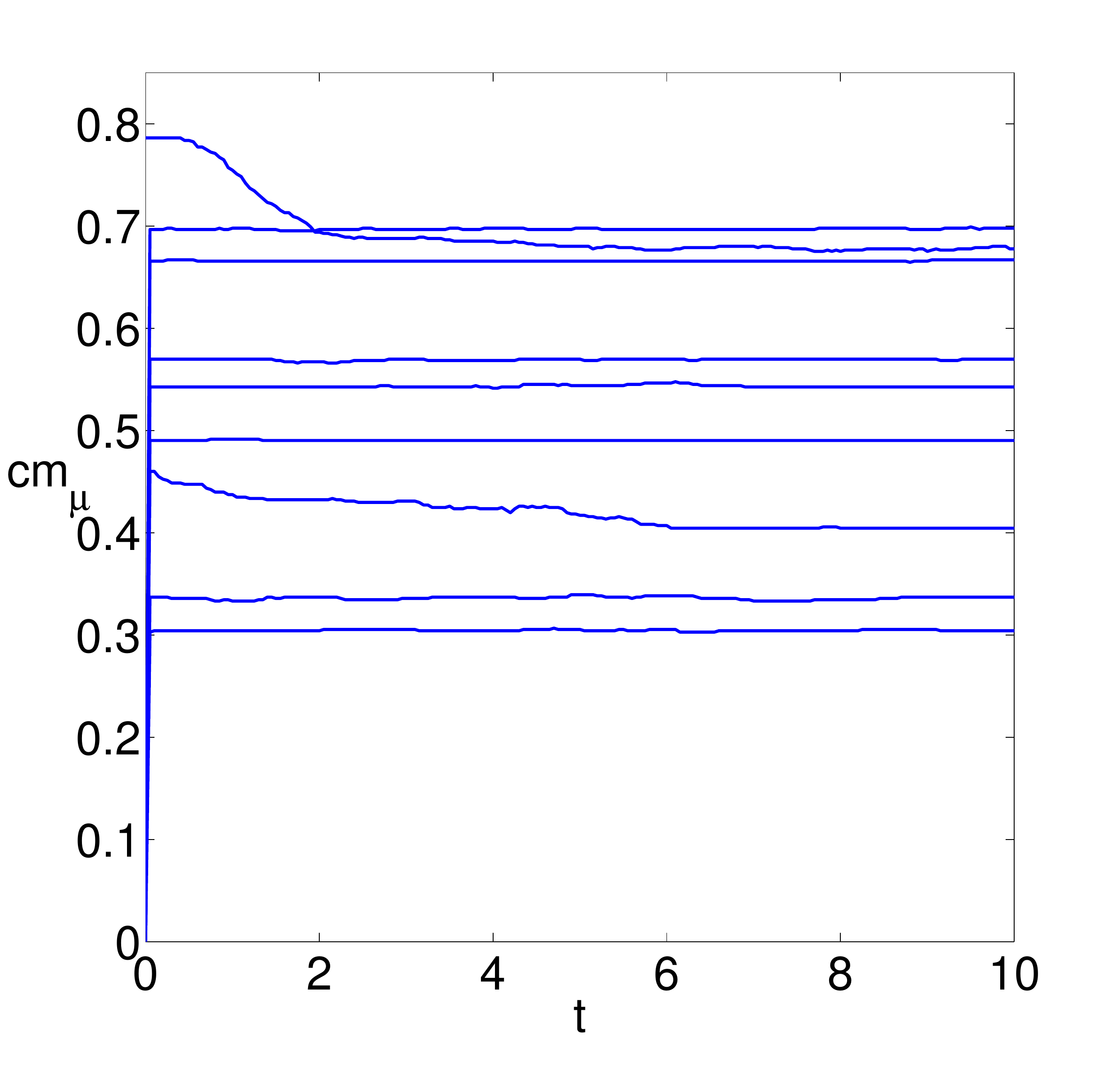}}
\caption{Monte Carlo simulations with $N=10^4$ spins. We plot the overlaps $m_{\mu}$ with different patterns $\xi_{\mu}$, as a function of time $t$, 
for $\hat{T}=0.015$, $c=0.8$, $\delta=0.25$, $\gamma=0.2$ and $\alpha=1$. }
\label{fig:hierar}\end{center}
\end{figure}
However, in this regime, it is more difficult to predict how the symmetry of patterns is broken, even at low temperature, 
due to the presence of a strong interference between patterns. In particular, it is not {\em a priori} clear whether a hierarchical retrieval 
involves all patterns or just a fraction of them. At zero temperature one may expect that a number of patterns will show 
an overlap $\order{(1)}$ with the system configuration. Assuming that the number of these patterns is $\order{(N^\gamma)}$ 
and the remaining (uncondensed) $\order{(N^\delta)}$ patterns have an overlap with the system configuration $\mathcal{O}(N^{\frac{\gamma-1}{2}})$,
we can write the noise distribution by splitting the contribution from the condensed (labelled by $\mu$) and the uncondensed (labelled by $\rho$) patterns
\begin{eqnarray}
P(z|{\bf m})= \int_{-\pi}^{+\pi}\frac{\rmd\hat{z}}{2\pi}\rme^{\rmi z\hat{z}
+\frac{c}{N^{\gamma}}\sum_{\mu}^{\psi N^{\gamma}}(\cos(\hat{z}m_{\mu})-1)} 
\rme^{-\frac{c}{2N^{\gamma}}\sum_{\rho}^{\alpha N^{\delta}}m_{\rho}^2\hat{z}^2}
\end{eqnarray}
where we have Taylor expanded the $m_{\rho}$-dependent terms for $m_{\rho}$ small.
When $N$ is large,
\begin{eqnarray}
P(z|{\bf m})\simeq P(z|\{m_{\mu}\})+\mathcal{O}(N^{\delta-1})
\end{eqnarray}
and the evolution of the uncondensed patterns is given, to leading orders, by
\begin{eqnarray}
\frac{\rmd m_{\rho}}{\rmd t}\simeq\int \rmd z P(z|\{m_{\mu}\})\tanh[\hat{\beta}(m_{\rho}+z)]-m_{\rho} + \mathcal{O}(N^{\delta-1})
\label{eq:unc}
\end{eqnarray}
which depends on the noise distribution of condensed patterns $\{m_\mu\}$ only.
Taylor expanding for small $m_{\rho}$ we obtain
\begin{eqnarray}
\frac{\rmd m_{\rho}}{\rmd t}\simeq m_{\rho} \bigg[ \hat{\beta}\int \rmd z P(z|\{m_{\mu}\})(1-\tanh^2(\hat{\beta}z))-1\bigg]+ \mathcal{O}(N^{\delta-1})
\end{eqnarray}
and for $T\to 0$ we get
\bea
\frac{dm_\rho}{dt}\simeq m_\rho[\hat{\beta} P(0|\{m_\mu\})-1].
\eea
As explained in \ref{sec:lowt2}, $P(0|\{m_\mu\})\geq e^{-\phi c}$ 
This suggests that the system will attempt to recall hierarchically all patterns, as supported by Monte-Carlo simulations shown in fig.~\ref{fig:hierar}.
However, the retrieval degrades with respect to the one described in 
(\ref{eq:hiergamma}), valid at the onset of pattern cross-talk, where the first pattern in the hierarchy was retrieved without errors.

\section{Conclusions} \label{sec:conclusion}

In this work we analysed the dynamics of associative memories with diluted patterns, in different regimes of pattern dilution and storage load 
(away from saturation).
We derived equations for the dynamics of a set of macroscopic order parameters, quantifying the simultaneous pattern retrieval of the system, 
from the microscopic evolution of the network dynamics. Results show that the network is able to accomplish a parallel retrieval of all the stored patterns,  
in the whole phase diagram, although the nature of the retrieval varies in different regions of the phase diagram. 
We analysed the structure and stability of the steady states solutions to the dynamical equations for the order parameters, for different ranges of 
the tunable parameters $T,\gamma, \delta$, controlling noise, dilution and storage load, respectively.

\begin{figure}
\centering
\scalebox{.4}{\includegraphics{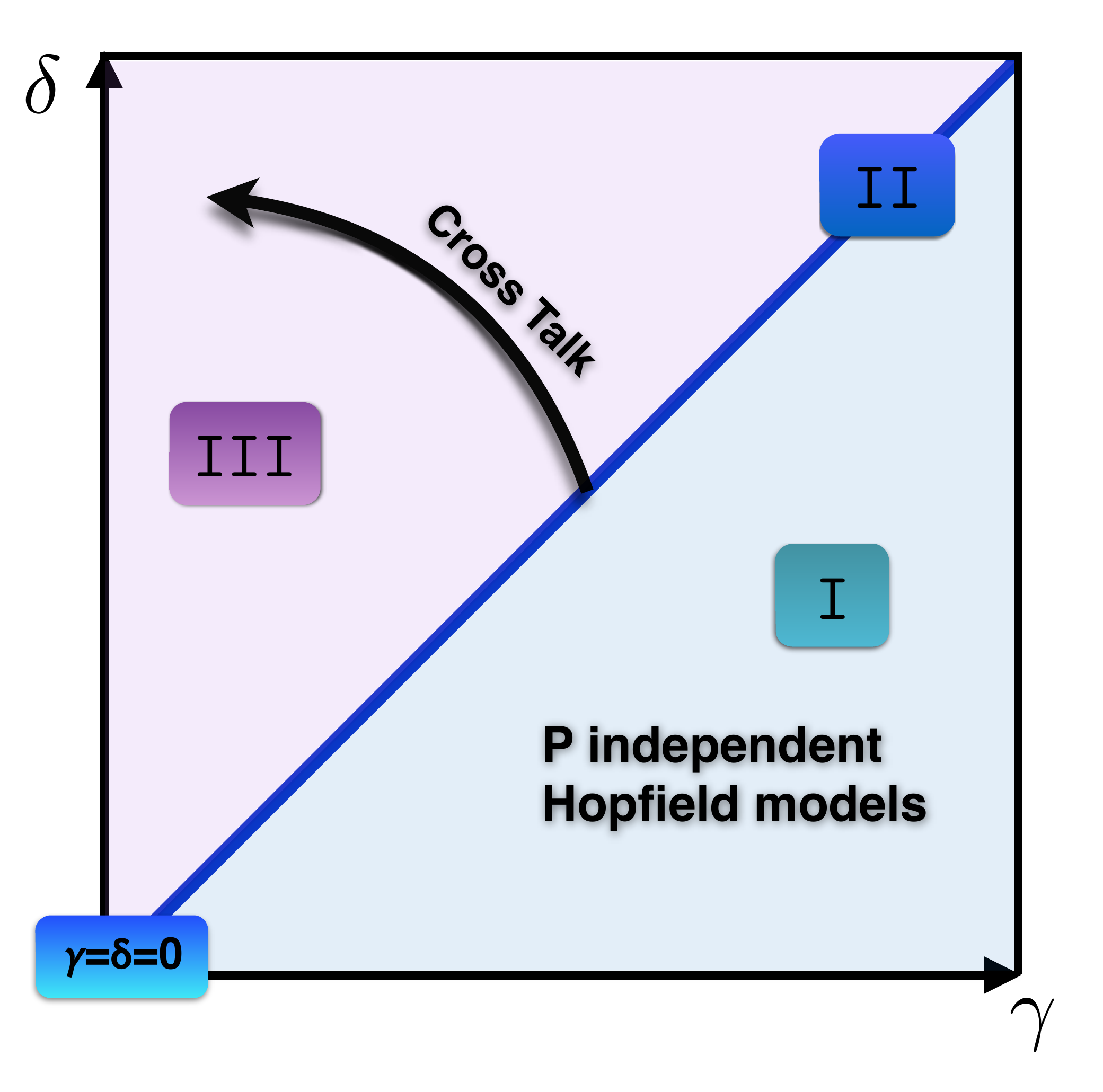}}
\caption{Schematic phase space $(\gamma,\delta)$. In region I $(\delta<\gamma)$, we have a set of $P$ independent ferromagnets and parallel retrieval is accomplished in a symmetrical fashion. In region II cross-talk between patterns appears; the intensity of symmetric recall is decreased due to the presence of a noise, whose distribution is identical to the one of a lazy, unbiased random walker, and  
the region of stability of the symmetric retrieval is provided in the phase diagram (fig. \ref{fig:phasediagram}).
A similar behaviour is found for $\delta=\gamma=0$, but the region of stability of the symmetric region shrinks (fig.~\ref{fig:critline2P}), due to the noise distribution retaining a pattern dependence 
(Sec. \ref{sec:noise}). Finally, in region III, cross-talk effects are strong and increase the larger $\delta$ and the smaller $\gamma$. These decrease the strength of symmetric retrieval  
via a Gaussian interference noise, and are seen to degrade gradually the hierarchical retrieval of the network, which thus retains its parallel processing capabilities.}
\label{fig:summary}
\end{figure}

We identified three regimes (see fig.~\ref{fig:summary}): (I) $\delta<\gamma$, (II) $\delta=\gamma$ and (III) $\delta>\gamma$.
In (I) there is no interference between patterns as these are relatively low in number and sufficiently diluted, so the network behaves as a set of 
$P$ independent ferromagnets each evolving according to the Curie-Weiss equation. As a result, the parallel retrieval accomplished by 
the system in this regime is symmetric for any temperature below criticality. In (II) a pattern cross-talk appears and symmetric 
mixtures are no longer stable for all temperatures below criticality. We derived the phase diagram showing their region of stability in the parameter space. 
The cross-talk between patterns manifests on symmetric mixtures as a noise with the distribution of a lazy 
random walker, taking discrete steps at discrete time (for $\gamma=0$) and in continuous time (for $\gamma>0$). The laziness 
of the random walker is due to pattern dilution and removes the periodicity in the return times, which is responsible for the different role played by 
even and odd mixtures, in the traditional (i.e. undiluted) Hopfield model, at zero temperature.
The region where symmetric mixtures are stable is broad for $\gamma>0$, where the 
noise distribution is pattern-independent, whereas the retrieval is mostly hierarchic for $\gamma=0$, where the symmetry of patterns is mostly 
a close-to-criticality effect. We rectify the result of previous analysis at zero temperature, which assumed symmetry of patterns, and 
provide a heuristic expression for the shape of the steady state solution at zero temperature, which finds very good 
agreement with numerical simulations.
Finally, in (III), the effect of cross-talk manifests on symmetric mixtures as a Gaussian noise, whose variance is directly related to the level of pattern interference, 
which decreases the quality of parallel retrieval. 
We derive the region of stability of symmetric mixtures and 
show that for large interference or low temperature, symmetric mixtures are unstable. In particular, at zero temperature the system is able to 
retrieve all patterns in a hierarchical fashion, despite the presence of strong interference may suggest that no retrieval is accomplished in this regime \cite{medium}.

 \section*{Acknowledgements}
We thank Prof. A. C. C. Coolen for useful discussions and helpful advice.

\end{document}